\pdfoutput=1
\documentclass[12pt]{iopart}

\usepackage{graphicx}
\usepackage{sidecap}
\usepackage[separate-uncertainty=true]{siunitx}
\newcommand{\ppf}{\emph{Physarum polycephalum}}
\newcommand{\pp}{\emph{P. polycephalum}}
\usepackage{xcite}
\usepackage[a4paper, total={7in, 10in}]{geometry}
\usepackage[margin=5pt,font=small,labelfont=bf]{caption}
\usepackage[usenames, dvipsnames]{color}

\usepackage[final]{pdfpages}
\makeatletter
\AtBeginDocument{\let\LS@rot\@undefined}
\makeatother

\usepackage{xr}
\newcommand*{\addFileDependency}[1]{
\typeout{(#1)}
\@addtofilelist{#1}
\IfFileExists{#1}{}{\typeout{No file #1.}}
}

\begin{document}

\title[Spatial mapping reveals multi-step pattern of wound healing ]{Spatial mapping reveals multi-step pattern of wound healing in \emph{Physarum polycephalum}}
\author{Felix B\"auerle$^{*}$, Mirna Kramar$^{*}$, Karen Alim}

\address{Max Planck Institute for Dynamics and Self-Organization, D-37077 G\"ottingen, Germany}
\address{$^*$Authors contributed equally to this work.}
\ead{karen.alim@ds.mpg.de}

\begin{abstract}
Wounding is a severe impairment of function, especially for an exposed organism like the network-forming true slime mould \ppf. The tubular network making up the organism's body plan is entirely interconnected and shares a common cytoplasm. Oscillatory contractions of the enclosing tube walls drive the shuttle streaming of the cytoplasm. Cytoplasmic flows underlie the reorganization of the network for example by movement toward attractive stimuli or away from repellants. Here, we follow the reorganization of \pp\ networks after severe wounding. Spatial mapping of the contraction changes in response to wounding reveal a multi-step pattern. Phases of increased activity alternate with cessation of contractions and stalling of flows, giving rise to coordinated transport and growth at the severing site. Overall, severing surprisingly acts like an attractive stimulus enabling healing of severed tubes. The reproducible cessation of contractions arising during this wound-healing response may open up new venues to investigate the biochemical wiring underlying \pp's complex behaviours.
\end{abstract}

\submitto{\JPD}
\maketitle

\twocolumn
\section{Introduction}
Simple organisms like fungi and slime moulds are able to display complex behaviours. This is surprising given that their network-like body plan lacks any central organizing centre. The slime mould \ppf\ has emerged as a model system to study the complex dynamics these organisms use to adapt to their environment. The organism has been shown to find the shortest path through a maze \cite{Nakagaki:2000} and connect food sources in an efficient and at the same time robust network comparable to man-made transport networks \cite{Tero:2010}. Furthermore, the slime mould distributes its body mass among several resources to obtain an optimal diet \cite{Dussutour:2010} and is able to anticipate recurring stimuli \cite{Saigusa:2008}.

\begin{figure*}[t]
\centering
\includegraphics[width=\textwidth]{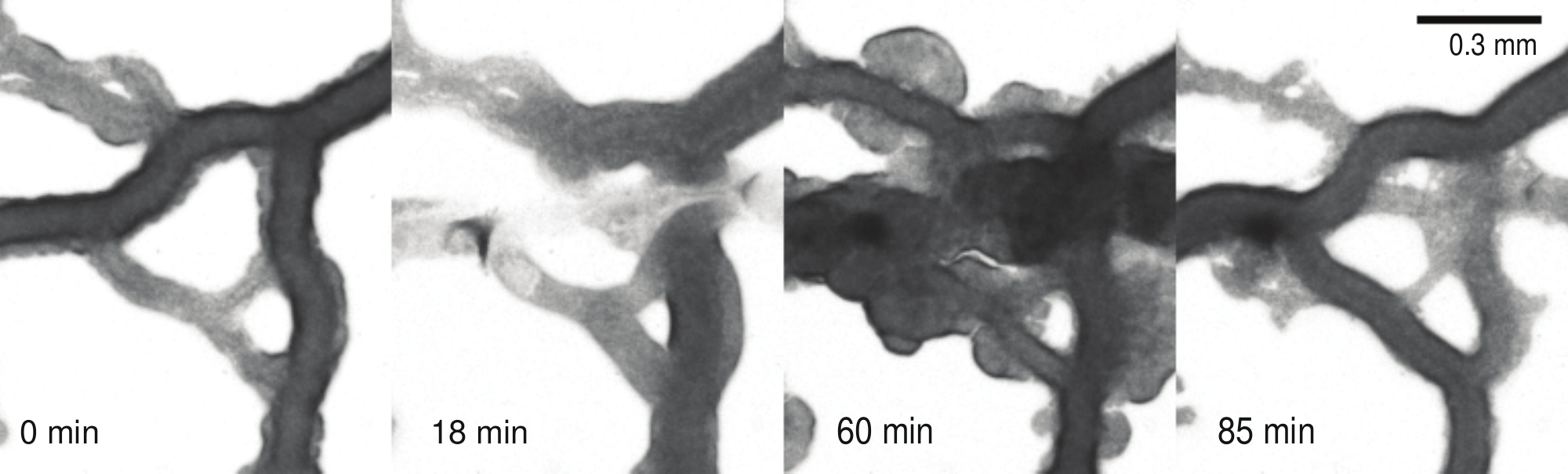}
\caption{Wound healing process in \pp\ illustrated at four time points using bright field images. The cut occurred at \SI{18}{\minute} and the fan grown at cut site reached its maximal size at \SI{60}{\minute}. The network morphology was restored after \SI{85}{\minute}.}
\label{img:cutsite}
\end{figure*}

\pp\ is a true slime mould that forms a plasmodial network. Nuclei keep on dividing without forming cell walls, which results in a syncytial web-like network. The cytoplasm within this tubular network flows back and forth in a shuttle flow \cite{Kamiya:1981}. These cytoplasmic flows are driven by cross-sectional contractions of the actin-myosin meshwork lining the gel-like tube walls \cite{WohlfarthBottermann:1979}. Flows are organized across the entire network in a peristaltic wave of contractions that matches organism size \cite{Alim:2013}. Flows generated in the organism are optimized for transport as contractions increase the effective dispersion of particles way beyond molecular diffusivity by a mechanism called Taylor dispersion \cite{Marbach:2016}.

\pp\ adapts its network-like morphology to its environment by chemotaxis \cite{Ueda:1976,DURHAM:1976,Chet:1977}. Here, stimulants are classified by being an attractant or a repellant depending on the organism's response to migrate toward or away from the stimulant. Stimulants have also been shown to affect cross-sectional contractions organism-wide by an increase in their frequency and amplitude for an attractant or a decrease for a repellant \cite{Miyake:1994,Hejnowicz:1980}. A variety of chemical stimuli have been discussed for \pp, with glucose being a prominent attractant and salts like NaCl being effective repellants \cite{Kincaid:1978,HIROSE:1982,MCCLORY:1985}. Temperature \cite{Matsumoto:1988,Takamatsu:2004} and light \cite{WohlfarthBottermann:1982,Nakagaki:1999} have also been found to act as stimulants that trigger organism-wide restructuring of the transport networks' morphology. In fact, the cytoplasmic flows themselves serve as the medium by which stimuli pervade the organism \cite{Alim:2017}. 

A lot less is known about the impact of mechanical perturbations on the organism. In its natural habitat the slime mould suffers predation from grazing invertebrates causing severing that disrupts the transport network and its cytoplasmic flows. In experiments it has been found that quickly stretching a strand to 10-20\% of its length while keeping it intact increases the amplitude of oscillations \cite{Kamiya:1972}. Excising a single strand from a plasmodial network has been observed to lead to a roughly 20 minute cessation of contractions in the strand until recovery \cite{Yoshimoto:1978}. This phenomenon was not observed for strands excised from the growing fan region of the slime mould resulting in speculations about the motive force being limited to the fan only. Yet, the cessation of contractions turned out to be hard to reproduce, see \cite{Cieslawska:1984} and references therein. Among these discordant observations what remains established is local gelation of cytoplasmic flows upon touch without severing the organism \cite{Achenbach:1981}. Despite the limited knowledge, wounding the organism by severing the network is part of daily laboratory routines and an eminent perturbation in natural habitat. 

Here we investigate \pp 's dynamics during wound healing following the quick and complete severing of a tube within the organism's network. We follow the process of wound healing across the individual's entire body, over the course of one hour after severing. The exemplary quantitative analysis of organism-wide contractions reveals a stepwise response spanning four different states. Briefly after severing, the contractions are often marked by an increase in amplitude and frequency, followed by a several minutes long cessation of contractions and stalling of cytoplasmic flows. This resting state is terminated by a sudden restart of vigorous contractions as the severed tube re-fuses. The vigorous state then transitions into a state of network-spanning contractions and continuous fan growth at the wounding site until the organism reverts back to pre-stimulus dynamics. Timing and significance of individual steps varies with the severity of cutting and cutting site location within the network. For example, stalling is found to be less pronounced when the network is cut in fan-like region. Overall, quick and complete severing triggers a response pattern with characteristics of the response to an attractive stimulus, including an increase in amplitude and frequency and net movement to stimulus site, see Fig.~\ref{img:cutsite}. The reproducibility of stalling clarifies earlier contradictions and at the same time opens new avenues to investigate the biochemical dynamics behind the highly coordinated acto-myosin contractions underlying \pp's arguably fascinating dynamics. 
\section{Methods}
\subsection{Culturing and data acquisition}
The plasmodium is prepared from microplasmodia grown in liquid medium. The recipe for the medium is inspired by \cite{Fessel:2012}, see Sec. S1. The advantage of this method over growing the plasmodium on oat flakes or bacteria is the ability to precisely control the nutritional state and amount of the organism. Also, plasmodia grown this way are free from oat flake residues or vacuoles containing food, which provides a cleaner sample for imaging. To prepare the plate for imaging, 0.2-0.5 mL of the microplasmodia grown in a shaking culture at $30^{\circ}$C are transferred to an 1.5\% agar plate and stored in a closed, but not sealed dish in the dark. After 12-24 hours, the microplasmodia fuse into a single plasmodium. The plasmodium is ready for imaging when there are no visible traces of liquid medium and the organism assumed its characteristic network shape, which usually occurs up to 36 hours after plating.

The imaging is performed with a Zeiss Axio Zoom V.16 microscope, equipped with a Zeiss PlanNeoFluar 1x/0.25 objective and a Hamamatsu ORCA-Flash 4.0 digital camera. A green filter (550/50nm) is placed over the transmission light source of the microscope to diminish \pp's response to the  light, and a humidity chamber prevents the sample from drying out. The acquisition of the images is done in Zeiss ZEN 2 (Blue Edition) software with bright-field setting. During the acquisition, the illumination of the sample is kept constant, and an image is taken every 3 seconds. The plasmodium is imaged for $\sim$1 hour before the application of the mechanical stimulus to allow for the accommodation to the light \cite{DURHAM:1976}. The stimulus is applied manually, using a microinjection needle with a blunt tip. The needle tip is held above the surface of the agar at a small angle and quickly dragged across the chosen plasmodial tube. The cut is severe and complete if the two parts of the tube separate completely. The plasmodium is then further imaged for more than 1 hour.

Using microplasmodia is so far the optimal way of obtaining non-severed networks, where the size and nutritional state are reproducible. However, there are challenges during the imaging that decrease the reproducibility of the experiment. In particular, plasmodia are highly motile and change their morphology accordingly. Furthermore, the organism tends to develop very large foraging fronts, which are not a suitable input for the presented comprehensive data analysis as they lack network characteristics. Lastly, the microscope light can act as stimulus \cite{WohlfarthBottermann:1982,Nakagaki:1999,Tero:2010}, and even the green-filtered low-intensity illumination may cause the network to respond and change its behaviour to escape the imaging region. These challenges combined make the reproducibility and required stability of the network morphology over time challenging.

\subsection{Comprehensive network-based contraction analysis}
To quantify contraction dynamics we analyse bright field recordings in two different ways: for two morphologically static networks  (see E2 and E3 in the experiment list) we perform an exhaustive network-based analysis as outlined in the following (see Fig.~\ref{img:05_lineplots} and Fig. S4). For the additional 19 specimen which alter their network morphology dramatically over the course of the experiment, we analyse kymographs along static parts of the network as described in detail in Sec. S3 (see exemplary E1 and Mov. S5).

\begin{figure}[t]
\centering
\includegraphics[width=0.5\textwidth]{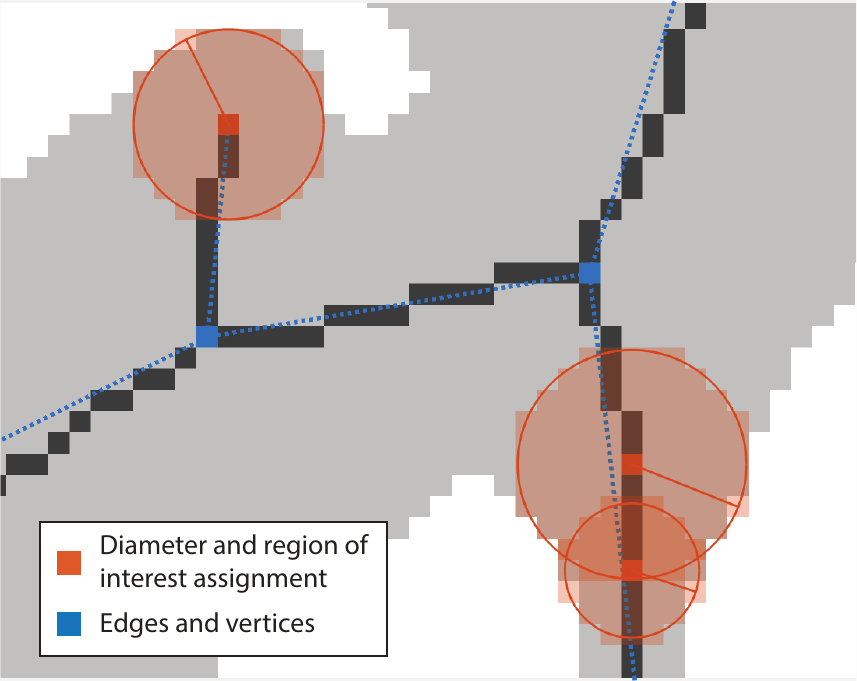}
\caption{Scheme of intensity and diameter data extraction based on \pp\ bright field images. The light grey area depicts the network mask based on the bright field images. Dark grey lines represent the network skeleton and the corresponding topology is shown in blue. Each pixel of the skeleton acts as a reference point for data derived during the analysis. The diameter is set as the distance from the reference point to the next non-mask pixel. The intensity is calculated by averaging individual pixel intensities over a corresponding disk (red).} \label{img:02_skel}
\end{figure}

Images recorded as a time series are processed as 8-bit uncompressed TIFs. At first every image is processed separately, then the results are stitched together, largely following Ref.~\cite{Alim:2013}, and lastly the collective is analysed. On every single image, background is removed with the rolling-ball method. Then the image is used to create a mask, a binary image, with an intensity threshold that separates the network from the background. The mask is enhanced further, i.e.~only the biggest structure is considered, small holes are filled and single-pixel edges are smoothed. Subsequently, the resulting mask is used as a template for extracting the network's skeleton with a thinning method. In the skeletonized mask each pixel can be understood as a data point representing local intensity and diameter (see Fig.~\ref{img:02_skel}). Local diameter is calculated as the largest fitting disk radius around the point within the mask. Within this disk the average intensity is computed and saved as intensity at the considered data point. Intensity and diameter anti-correlate due to the optical density of the slime mould and can therefore be used interchangeably considering Beer-Lambert law. Individual data points are attributed to a specific network branch of the network skeleton. To represent network topology, the network is broken down into vertices and edges where vertices describe pixel positions of branching points and edges represent two connected vertices. Each edge then acts as a parent for one specific branch. In this sense edges are abstracted simple connections and branches represent pixel-based resolution of a tube.

After the network is extracted in space, the edges, vertices, diameters, and intensities are concatenated in time. To map intensity and diameter over time, a reference image is used, usually from an early time point. For every data point the shortest distance to any pixel in the reference image is calculated. This gives a quasi-static (x, y, t) $\rightarrow$ (intensity, diameter) dataset, i.e.~the topology and vertex positions stay the same, but  intensity and diameter can vary. This is justified as long as growth of the organism and vertex movement is minimal. The oscillatory behaviour of tubes in a certain time window can be described by four time dependent variables, namely amplitude $A$, frequency $f$ (or period $P$), phase $\varphi$ and trend (base diameter) $d$. Each can be calculated from the time-evolution of the diameter or the intensity data, but if not stated otherwise the following results are only derived from intensity analysis. 

\begin{figure}[h]
\centering
\includegraphics[width=0.47\textwidth]{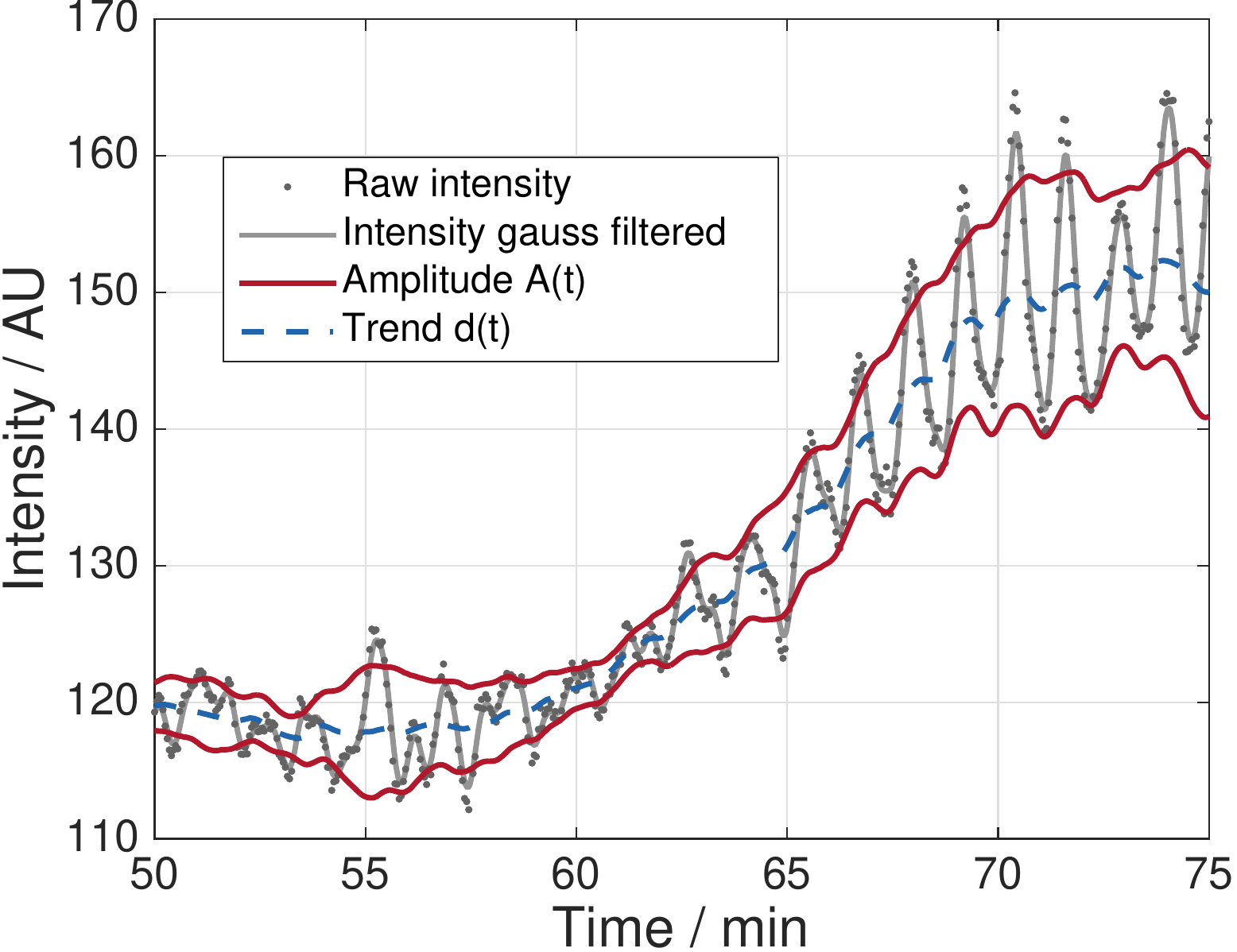}
\caption{Derivation of oscillation specific parameters, i.e.~amplitude $A$(t), frequency $f$(t) and trend $d$(t), from single pixel time series. The trend is calculated using a moving average with a kernel width of \SI{200}{s}. Intensity is filtered with a Gaussian of width \SI{39}{s}.  Amplitude and frequency are calculated from the absolute value and angle of the complex-valued analytic signal, respectively.} \label{img:03_linecomp}
\end{figure}

The trend $d(t)$ is obtained with a moving-average filter with a kernel width of \SI{200}{s} on each time trace (see Fig.~\ref{img:03_linecomp}). The dataset is detrended with the calculated trend and smoothed with a Gaussian using a kernel width of \SI{39}{s}. The kernel widths were chosen to extract the characteristic contraction pattern which usually has a frequency of \SI{\sim90}{\second}. The values at every data point are stored as a complex valued time array, with the detrended and smoothed intensity representing the real part and the corresponding Hilbert transform representing the complex part, see S2 for more details. This time array, denoted analytic signal, serves as a basis to get instantaneous phase, frequency and amplitude by computing the angle or absolute value of the complex time series. Finally, the results are mapped back onto the network structure for each time point. In this fashion one can follow oscillatory behaviour resolved in time and space. Furthermore, the maps can be clustered in sub-networks and averaged separately to pinpoint local events in time. It should be mentioned that averaging of results for line plots, i.e.~Fig.~\ref{img:05_lineplots}, is always done after the data-point based analysis took place. In this way for example, the apparent amplitude of the averaged intensity (Fig.~\ref{img:05_lineplots}D) can be lower than the amplitude of each data point averaged (Fig.~\ref{img:05_lineplots}B).
%
\section{Results}
\begin{figure*}[h]
\centering
\includegraphics[width=\textwidth]{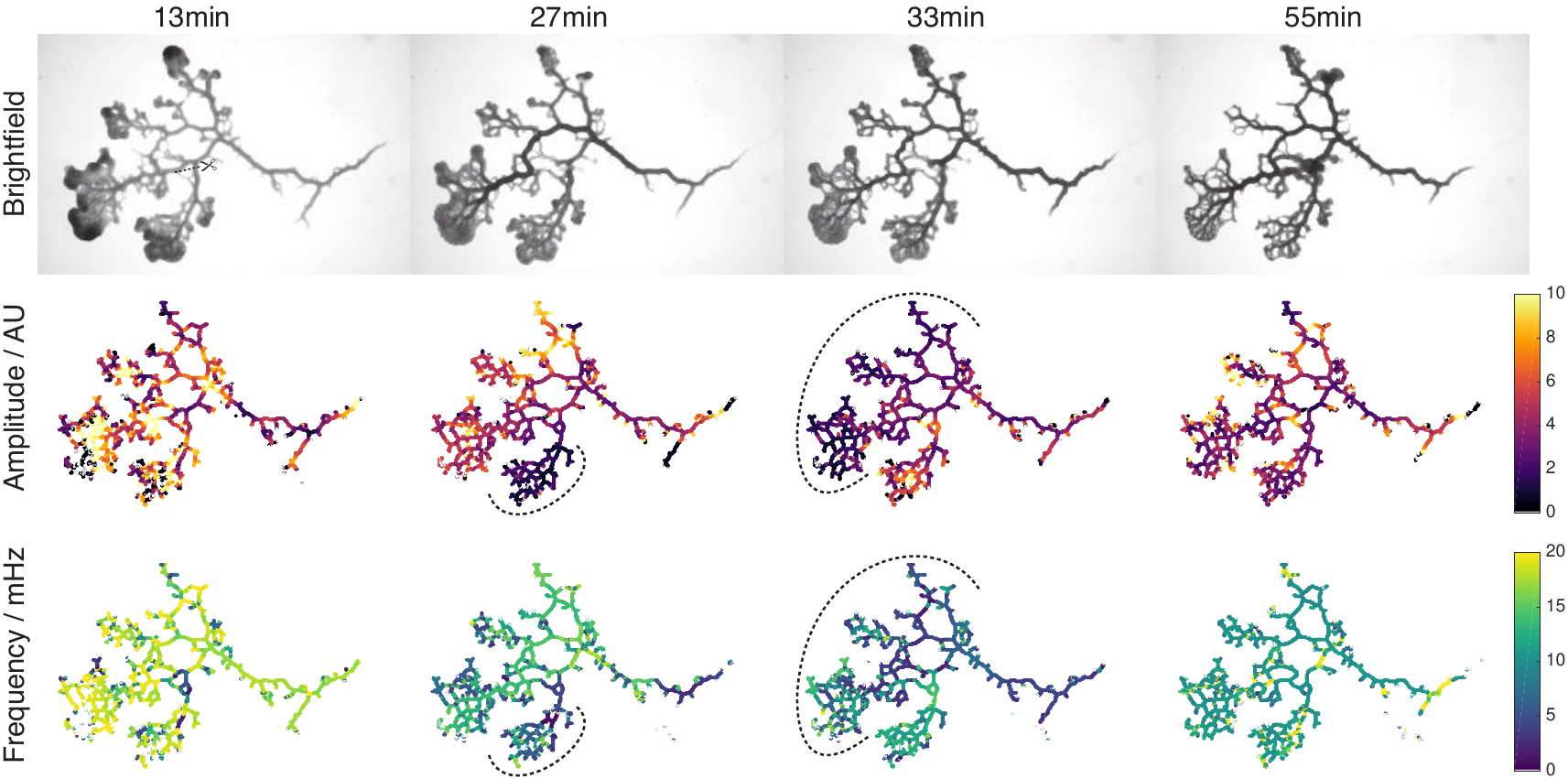}
\caption{Time evolution of an exemplary network and its spatially mapped oscillation parameters at \SI{13}{\minute}, \SI{27}{\minute}, \SI{33}{\minute} and \SI{55}{\minute}. The network was cut in the centre at \SI{17.3}{\minute} (\emph{scissor icon}). Top row depicts the raw bright field data, middle row  the local amplitude, and bottom row the local frequency. Amplitude and frequency decrease locally, first at the lower sub-network (\emph{small dotted arc}) at \SI{27}{\minute}, subsequently at upper sub-network (\emph{large dotted arc}) at \SI{33}{\minute}.  At \SI{38}{\minute} cytoplasmic flows are re-established at the wounding site. Finally, amplitude and frequency values recover.} \label{img:04_maps}
\end{figure*}
\subsection{Wounding induces fan growth at cut site}
We observe specimens before and after a quick and complete severing of a tube to follow the response of \pp\ to wounding (see Fig.~\ref{img:04_maps}A, Mov. S1 and Mov. S5). Bright field movies reveal that cutting of main tubes distal to fans triggers cessation of contractions followed by stalling of cytoplasmic flow (n=15 out of 21). After contractions resume the severed tube fuses back together (n=21 out of 21), i.e.~flow is re-established, and a fan starts to grow at the cut site.  Furthermore, we observe accumulation of body mass close to the cut site which is most prominent in peripheral cuts (Fig. S2). However, the growth is transient and after a given time the initial morphology is restored and the organism returns to typical behaviour comparable to before wounding.\\

In consideration of previously mentioned technical limits, we selected one representative dataset with prominent discernible features for network-based analysis. The following findings are derived from this dataset and later compared with other experiments. The specific timing of events in the representative data set is as follows (see Fig.~\ref{img:04_maps}). Two tubes are severed at \SI{17.3}{\minute} effectively dividing the network into two parts. In both sub-networks, the size-wise bigger and smaller part, flows stall transiently around \SI{30}{\minute}. At \SI{38}{\minute} a connecting tube is reinstated and starts to re-establish cytoplasmic flows across the cut site. Until about \SI{63}{\minute} a transient fan is created at the cut site. At \SI{90}{\minute} the initial morphology is restored and fans are grown elsewhere.

\subsection{Spatial mapping reveals localized stalling}
We perform network-based analysis on the wounded specimen to extract the interplay of contractions during the healing response. In particular, we map out the amplitude and frequency of contractions spatially (see Fig.~\ref{img:04_maps}, Mov. S2 and Mov. S3). This allows us to exactly localize the onset of stalling as it goes hand in hand with low values of amplitude and frequency. Likewise, patterns in contraction dynamics in a region of interest are identified by spatially averaging amplitude and frequency in this region (see Fig.~\ref{img:05_lineplots}).

In the representative dataset, wounding separates the network into two sub-networks. Spatial mapping reveals that oscillations cease on different time-scales in the two sub-networks. By identifying the two sub-networks as separate regions of interest, we quantify the patterns in contraction taking the spatial average of the respective contraction variables in each region. The small sub-network shows a drop in amplitude at \SI{21.5}{\minute} by \SI{63}{\percent} and only recovers eight and a half minutes later to comparable values. Here, the percentage is given as ratio of time averages before, during and after stalling. In detail, the averages of the first \si{21.5} minutes, the \si{9.5} minutes during stalling and \si{15} minutes after stalling were considered. The bigger sub-network drops significantly later at \SI{28}{\minute} by \SI{51}{\percent} and recovers to \SI{29}{\percent} below the initial value nine minutes later. In the same time frames the frequency drops by \SI{32}{\percent} and \SI{45}{\percent} for the small and big sub-network, respectively. Yet, neither sub-network recovers its frequency fully right after the stalling phase. Only the small sub-network recovers 35 minutes later to initial frequencies whereas the bigger region levels off \SI{35}{\percent} below the initial value.\\
Furthermore, the phase patterns over time (see Mov. S4) reveal changes in the travelling waves upon cutting. Initially (0 to \SI{17.3}{\minute}) one can observe peristaltic waves from the tail (right-hand side) to the front (left-hand side) which finally merge into concentric patterns in the fan regions. Then, at 18 to \SI{30}{\minute}, the small sub-network slows down noticeably (see change in frequency) and the big sub-network contracts with less apparent spatial correlation, i.e.~the peristaltic wave pattern is temporarily lost.

\subsection{Fan growth phase coincides with stable network-spanning contractions}
After re-fusing of the two sub-networks, another distinct phase characterized by stable network-spanning contraction dynamics can be observed. In Fig.~\ref{img:05_lineplots}D contractions appear uniform from \SI{44}{\minute} until \SI{63}{\minute}. During this phase, amplitude and frequency level off to a stable value with little fluctuations. The small sub-network shows a slight increase in frequency over this period and has more fluctuations in the average intensity data than the big sub-network. Note, that the time frame of these contractions coincide with fan formation at the cut site. Furthermore, the end of this phase also coincides with the largest fan in respect to area.\\
Network-spanning contractions are further supported by the phase time series. When considering the phase development one can already observe a peristaltic wave travelling towards the cut site in the small sub-network as early as \SI{30}{\minute}. A spanning pattern in the large sub-network is reinstated around the \SI{35}{\minute} mark and a global pattern (small and large sub-network) appears roughly three minutes after re-fusing (\SI{40}{\minute}). Then a standing wave pattern appears between the central region including the cut site and the periphery. It is stable and network-spanning until \SI{63}{\minute}. Subsequently the phase pattern breaks into a peristaltic wave similar to pre-cut and propagates from the tail and the small sub-network into fan regions in the large sub-network.

\subsection{Stalling and fan growth periods are bridged by distinct transition periods}
\begin{figure}[hpt]
\centering
\includegraphics[width=0.47\textwidth]{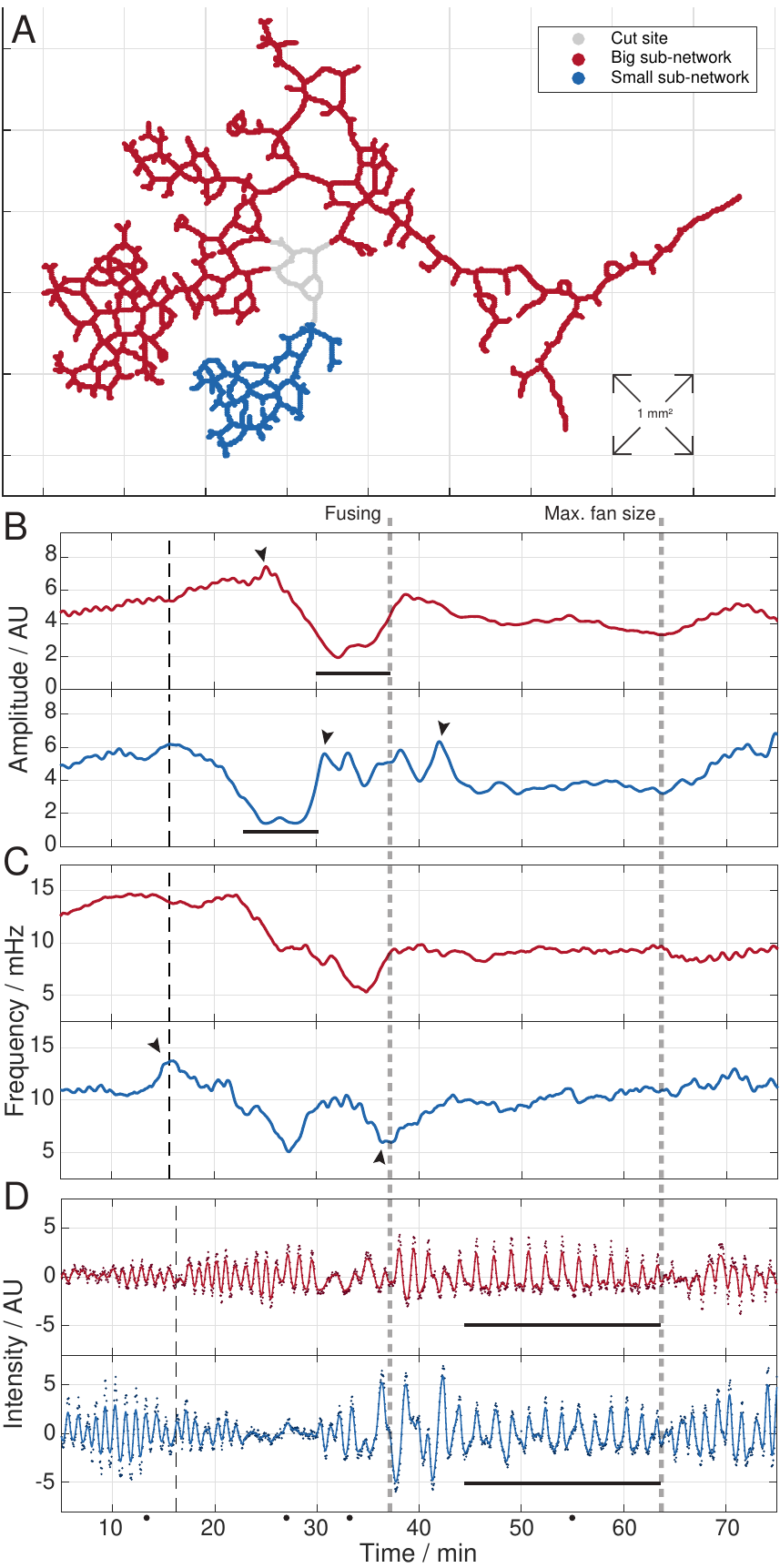}
\caption{Comparison of oscillation parameters in the big- and small-sub-network depicted in (A) which result from the cut. Grey area (cut site) is not considered in the analysis. Time series of amplitude (B), frequency (C) and intensity (D) are averaged in the respective domains and compared; top : big sub-network, bottom: small sub-network. In each of these plots the black dashed line indicates the moment of the cut. The first grey dashed line marks the time point of fusion and the second the moment of maximal fan size. Black bars underline periods of stalling in (B) and fan growth in (D). In (D) the solid line represents the Gaussian filtered intensity (kernel width = 39s) and markers show raw averaged data. Black arrows indicate respective extremal peaks in the transition periods. Four black dots on the time line correspond to the four time points chosen in Fig.\ref{img:04_maps}.
}\label{img:05_lineplots}
\end{figure}

Closer analysis of contraction dynamics over time reveals that the time point of the cut, the stalling phase and the fan growth phase are transitioned by phases of high fluctuations. Particularly in the presented case, before stalling occurs, amplitude and frequency peak shortly in both sub-networks (see arrows in Fig.~\ref{img:05_lineplots}). In the small sub-network this peak coincides with the cut, whereas another ten minutes pass for the big sub-network before the amplitude reaches its maximum. Surprisingly, here the frequency decline occurs three minutes before the amplitude drops. After stalling the amplitude increases sharply in both sub-networks, yet stays below previous values in the big sub-network. The small network undergoes a phase of roughly \SI{13}{\minute} where the amplitude oscillates vigorously. This also coincides with a second frequency drop even though there is no apparent drop in amplitude at this time point. After the fan growth phase, amplitude and frequency show slight gradients once more. Here behaviour becomes comparable to the pre-cut state as the slime mould develops a preferred growth direction in the periphery and continues foraging.

\subsection{Fan creation and stalling is reproducible for complete severing}
For comparison we analysed a second dataset with the same network-based method (see Fig. S4). The key features, i.e.~cut repair, stalling, a transition phase, stable network-spanning contractions and return to pre-cut behaviour are found likewise, but the succession and timing of the specific events vary. This dataset has a weaker fan growth at the cut site and the time point of maximal fan size follows immediately after fusion. Given the short period of fan growth global network-spanning contractions are not observed. However, standing phase wave patterns  are visible in the larger sub-network before fusion. Lastly, the transition phase shows peaking amplitude and fluctuating frequencies and reverberates for more than 30 minutes. At \SI{70}{\minute} the network reinstates a peristaltic wave toward peripheral fan regions resuming pre-cut dynamics.\\

In further experiments analysed with a kymograph based approach, we confirmed stalling to be a common response after a cut (see Fig. E1, n=15 out of 21). However, the degree and duration of stalling is varying between experiments and is most reproducible for a severe cut close to the centre of the network.\\
In detail, we observe that both the degree and duration of stalling, depend on the network size and morphology, cut location, possibilities of re-routing the flow through neighbouring tubes and presence of large fans. Also, a network undergoing quick changes in morphology due to a presumed light shock is less likely to show stalling. Varying cut location shows that complete severing of a tube, with a diameter comparably large in size and few neighbouring tubes, results in strong stalling, see experiments E[2, 3, 5, 6, 8, 9, 12, 13, and 18]. The effect is even more pronounced in smaller networks and on tubes close to the centre of the network (E[2, 3, 5, 8, and 18]). Stalling is less pronounced, as measured by relative change in amplitude and frequency as well as visual inspection of bright field data, if severing was applied to fan-like regions or peripheral tubes (E[10, 11, 14, 15, 16, 17, 19, 20, and 21]). If a severed tube had alternative routes with a comparable flow direction, neighbouring tubes inflated shortly after the cut, indicating a re-routing of flow. Yet, in this case stalling severity ranged from non-existent (E19) to full-stop (E1). In all data sets fan growth is observed around the cut site, yet duration and fan sized varied greatly (see E2 and E9 as maximal and minimal examples). \\
In all 15 experiments that show stalling, the period lasted for a minimum of three minutes. The exact time point of stalling onset and its duration varied. Duration of transition periods also varied from complete omission up to \num{22} minutes between cut and stalling. In 7 out of 15 experiments, a vigorous phase of increase in frequency or amplitude fluctuations could be observed in the transition phases.

\section{Discussion}
We investigate \pp's response to wounding in the form of a quick and complete severing of tubes using bright field microscopy and quantitative analysis of contraction patterns. Mapping out the contractions amplitude and frequency in space and time allows us to uncover a multi-step pattern of wound healing in \pp. 

The key of our network-based analysis is mapping contraction variables onto a few pixels serving as the skeletonized backbone of the complete network. This representation allows us to capture contraction dynamics across the entire network over the course of several hours with handleable amount of data. Furthermore, spatial mapping visualizes abstract variables in an approachable way which outlines region of interests or patterns in space. For example, in the representative data set the time-shift in the response pattern between the two sub-domains of the network would have been lost when averaging contraction dynamics across the entire network (see Fig. S1).

Among the multiple steps in the response to wounding the cessation of oscillations and stalling of the cytoplasmic streaming is most striking. The phenomenon of stalling of cytoplasmic flows has been observed previously ~\cite{Kamiya:1972,Yoshimoto:1978}, but its reproducibility was deemed questionable \cite{Cieslawska:1984}. Our work shows that cut location and severity are crucial parameters for inducing reproducible stalling. The stalling period is omitted when a tube is not completely severed, or cut in a way that allows the cut ends to rejoin quickly. In addition, the specific body plan affects the impact of a cutting stimulus. For example, severed fan-like regions show less pronounced stalling. However, we find reproducible strong stalling in networks where the affected tubes are crucial connections that cannot be re-routed easily - thereby clarifying previously discordant observations.\\

Stimuli are commonly classified into attractants or repellants. The response of \pp\ to an attractive stimulus includes fan growth and mass transport towards the stimulus site, often accompanied with an increase of oscillation frequency and amplitude. When we apply a wounding stimulus resulting in complete cutting of a tube, we observe a multi-step response pattern where only two out of four steps show a noticeable increase in amplitude and frequency. Yet, wounding implies that the network architecture is perturbed. Taken into account that contraction frequency decreases as organism size decreases \cite{Kuroda:2015} the impairment of network architecture itself might counteract any increase in frequency. Despite the weak indication from contraction frequency and amplitude, we always observe fan growth and movement of mass toward the cut site regardless of the tube hierarchy, plasmodium size or the severity of the cut. Fan growth is a lot bigger than initial spillage of cytoplasm due to cutting. Furthermore, we often identify a specific fan growth phase of network-spanning contractions well separated in time from the cutting event by the stalling phase. We therefore identify wounding as an attractive stimulus. The observation of network-spanning oscillations during fan outgrowth adds to our confidence about cutting being an attractive stimulus since the observed phase patterns resemble contraction patterns found in earlier work with attractive stimuli using glucose as a stimulant \cite{Alim:2017}.\\

Employing spatial data analysis we uncovered that wounding triggers a choreography of multiple successive steps to heal the severed tube. The mere duration of the healing response now defines a suggested minimal wait time after trimming for \pp\ experiments. The complexity of the response hints at an intricate signalling pattern underlying the coordination of contractions. It is likely that also the response to classical attractants and repellants, when scrutinized, reveal multiple steps. Unravelling the workings behind \pp's ability to adapt, is arguably a fascinating albeit challenging question. Here, the reproducible cessation of contractions arising during this wound-healing response may open up new avenues to investigate the biochemical wiring underlying \pp's complex behaviours. Furthermore, it is fascinating that the impact of wounding can be weakened by network architecture. This suggests that \pp's body plan itself could be part of the organisms strategy to not only adapt to its environment, but also specifically prevent severe consequences of wounding. 

\section*{Acknowledgements}
We thank Christian Westendorf for instructions on growing microplasmodia, as well as for invaluable discussions and advice. M.K. and F.B. acknowledge support by IMPRS for Physics of Biological and Complex Systems.

\section*{Bibliography}
\bibliographystyle{iopart-num}
\bibliography{cutnrelax.bib}

\pagebreak
\noindent


\section*{Supplementary material}

# Supplemental Information: Spatial mapping reveals multi-step pattern of wound healing in *Physarum polycephalum*

**Felix Bäuerle**∗, **Mirna Kramar**∗, **Karen Alim**

Max Planck Institute for Dynamics and Self-Organization, D-37077 Göttingen, Germany

∗ Authors contributed equally to this work.

E-mail: `karen.alim@ds.mpg.de`

## S1. Supplemental Text: Preparation of microplasmodia and buffer recipes

The microplasmodia for the shaking culture are grown from a healthy network feeding on heat-killed HB101 bacteria. A section of network is transferred together with the agar slice into a new plate without food. Once the network crawls onto the agar in the new plate, the agar around the network is excised from the dish and the dish is filled with the a 1:1 mixture of SDM and BSS medium up to the agar level, resulting in an agar island with the network on it.

The dish is left to incubate for several (2-4) days until the plasmodium starts to form a network in the liquid medium. The floating network is gently transferred to a flask containing 15 ml of SDM and 25 ml of BSS and kept in a shaking culture (150 rpm), where it breaks into microplasmodia.

The microplasmodia need to be resuspended every 2-3 days. 3 ml of microplasmodial culture is aspirated with a pipette, transferred to a new flask with 25 ml SDM and 25 ml BSS, and stored in the shaker.

### S1.1. Preparation of the SDM buffer

For 1 l of SDM buffer, dissolve 10 g of glucose, 10 g peptone from soybean, 3.54 g citric acid monohydrate, 2 g $KH_2PO_4$, 1.026 g $CaCl_2$x$2H_2O$, 0.6 g $MgSO_4$x$7H_2O$,0.0034 g $ZnSO_4$x$7H_2O$, 0.0042 g thiamine-HCl, and 0.0016 g biotin in 0.8 l of water. Adjust the pH to 4.6 with KOH, fill the rest of the water and autoclave for 20 20 min. Store at room temperature. Before use, add 10 ml of hemin solution.

### S1.2. Preparation of the BSS buffer

For 1 l of BSS buffer, dissolve 3 g of citric acid monohydrate, 4.20 g $K_2HPO_4$x$3H_2O$, 0.25 g NaCl, 0.21 g $MgSO_4$x$7H_2O$,0.05 g $CaCl_2$x$2H_2O$ in 0.8 l of water. Adjust the pH

to 5.0 with KOH, fill the rest of the water and autoclave for 20 min. Store at room temperature.

## S2. Supplemental Text: Hilbert transformation

This supplemental information is based on [1]. To obtain the quantitative key parameters namely amplitude $A$, frequency $f$ and phase $\varphi$ of contractions, we use the Hilbert transformation.

The Hilbert transform $h(t)$ of a function $a(t)$ is defined as:

$$h(t) = \frac{1}{\pi} \int_\infty^\infty \mathrm{d}\tau \frac{a(\tau)}{t - \tau} \tag{S1}$$

where $h(t)$ is phase shifted by $-90°$ for all positive frequencies and by $90°$ for all negative frequencies. With the real function $a(t)$ and its Hilbert transform $h(t)$ one can form the complex analytical signal

$$c(t) = a(t) + ih(t), \tag{S2}$$

which can be used to derive the instantaneous amplitude

$$A(t) = |c(t)| \tag{S3}$$

and phase

$$\varphi(t) = \arg(c(t)). \tag{S4}$$

The frequency is given by differentiating the phase $f = \dot{\varphi}$.

For the presented results each data point (skeleton pixel) is treated as a discrete time series which can be Hilbert transformed accordingly into an analytical signal. To approximate the analytic signal from discrete data, we calculate the fast-fourier transform (FFT) of the input sequence, replace those FFT coefficients that correspond to negative frequencies with zeros, and calculate the inverse FFT of the result [2].

The obtained key parameters are stored as a set for each pixel, i.e. $[a(t), A(t), \varphi(t), f(t)]_k$ where $k$ represents all pixels in space.

## S3. Supplemental Text: Kymograph based analysis

As described in section 2.1 the control of plasmodia network parameters like fan amount, motility, size or grid density is limited mostly to selection of parallel culturing. Therefore quantitative analysis as described in 2.2 is not suitable for every experiment. To analyse the bulk of experiments we use a kymograph based approach described in the following. A kymograph is a 3D representation of designated data with one space coordinate, one time coordinate and one value (intensity) coordinate. In our case space is represented as line through a main vein in a network (compare with E1, red line). Along this line a 3px orthogonal average of the light intensity is taken for the value coordinate. This is done for every image and concatenated in time.

Along the time coordinate the Hilbert transform is performed for every point in space

(see S2). Here, additionally, the amplitude is computed with a root-mean-square moving average (RMS-MA) filter besides the absolute value of the analytical signal. After computation the results are averaged in space and plotted in a line plot to depict contraction dynamics.

## S4. Supplemental Figures

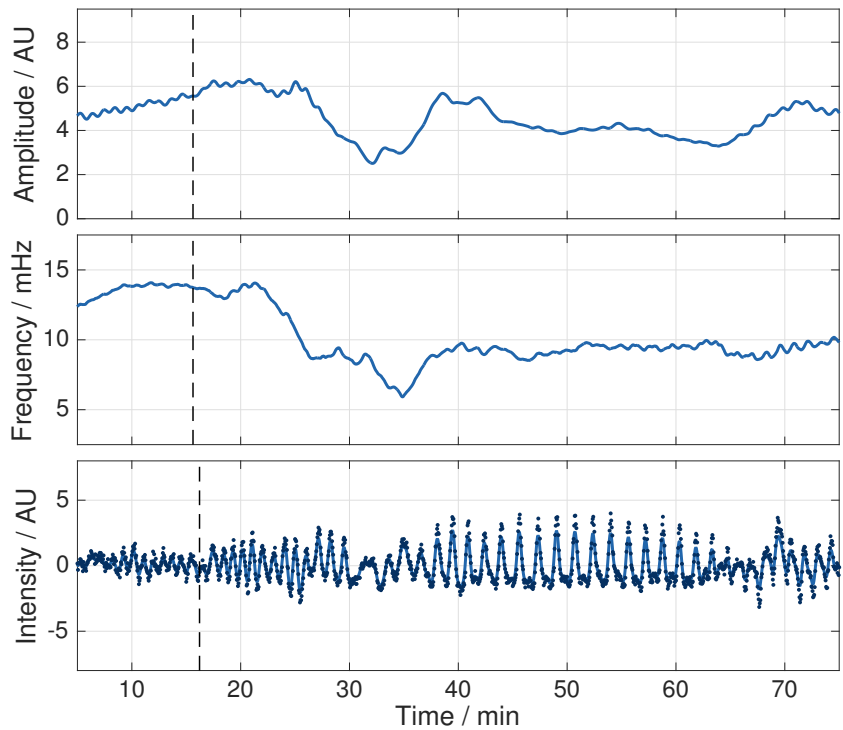

**Figure S1.** Analysis of averaged data from the whole network presented in the main text. The dotted line indicates the cutting event. Dotted points in the intensity plot show raw data and the line depicts Gaussian filtered data. Dynamics of small sub-network are lost in comparison to Fig. 5.

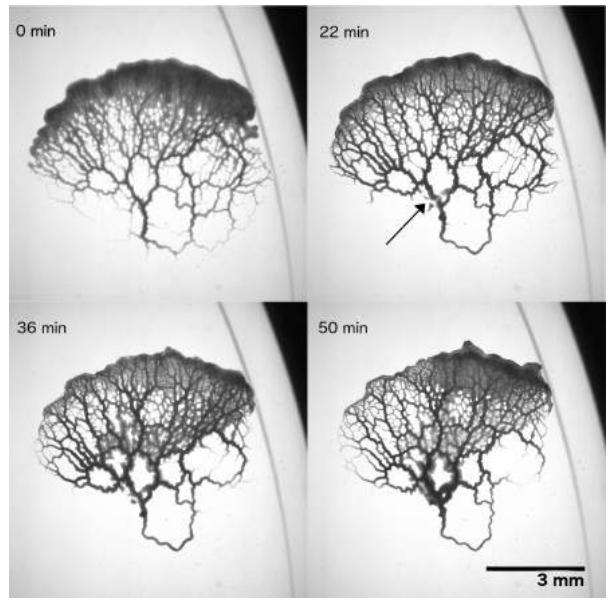

**Figure S2.** Material is transported towards the cut site (arrow). In the minutes after the cut, body mass accumulates along the general direction of the cut site and fans develop around it.

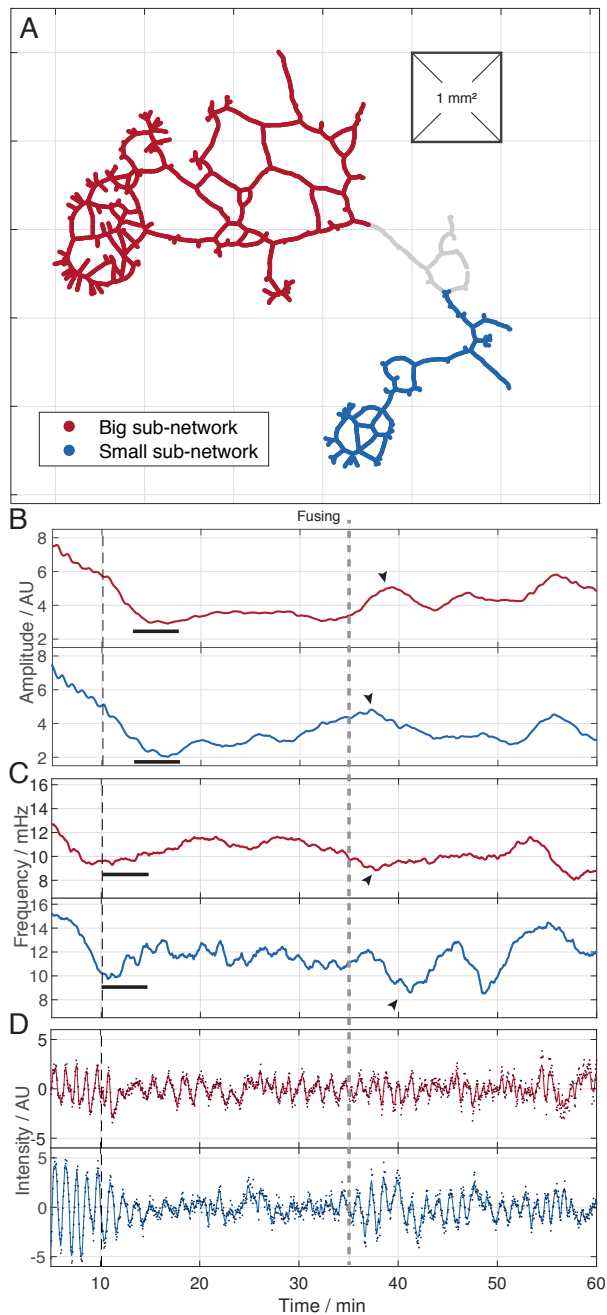

**Figure S3.** Full quantitative analysis of a second dataset (E3) for comparison. Oscillation parameters in the big- and small-sub-network depicted in (A) which result from the cut. Grey area (cut site) is not considered in the analysis. Time series of amplitude (B), frequency (C) and intensity (D) are averaged in the respective domains and compared; top : big sub-network, bottom: small sub-network. In each of these plots the black dashed line indicates the moment of the cut. The grey dashed line marks the time point of fusion. Black bars underline periods of stalling in (B) and (C). In (D) the red and blue solid lines represents the Gaussian filtered intensity (kernel width = 39s) and markers show raw averaged data. Black arrows indicate respective extremal peaks in the transition periods.

In comparison to Fig. 5 stalling is not as pronounced in frequency and occurs simultaneously in both sub-networks. Neither frequency or amplitude recover fully, but both are already declining pre-stimulus. The transition phase is visible after fusing with peaks in amplitude and decline in frequency. Network-spanning contractions are occurring in the big sub-network before fusion and are directed towards fan regions and the cut site. In general, the fan size at the cut site is comparatively small and reached maximal size shortly after fusion.

## S5. Supplemental Movies

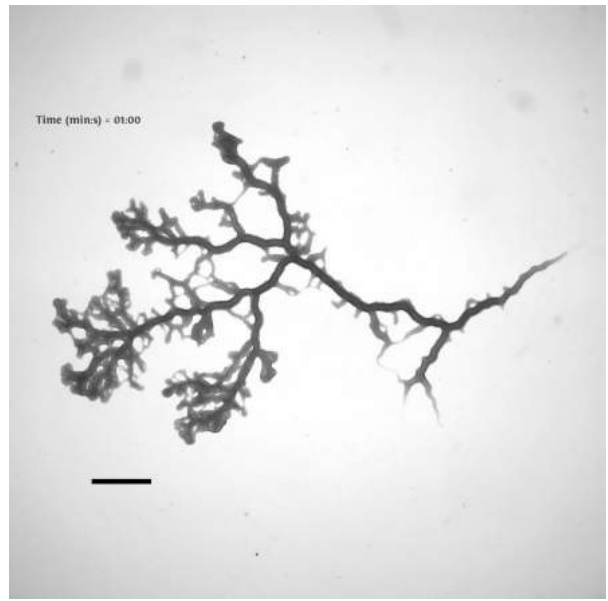

**Movie S1.** Bright field movie of the representative data set for wound healing. Note, that the cessation of oscillations in the top and bottom part (big and small sub-network) is on different time scales. Scale bar = 1mm.

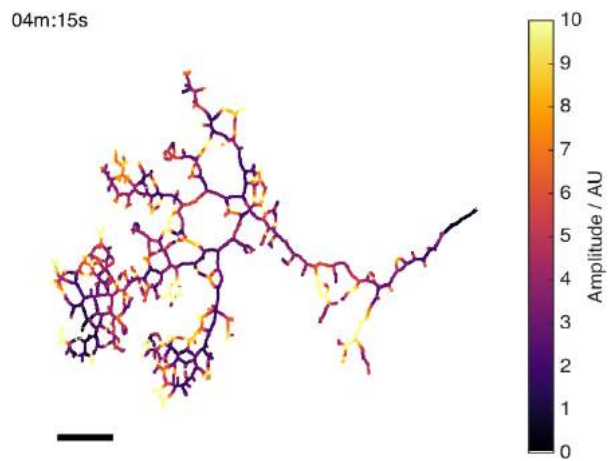

**Movie S2.** Movie of the oscillation amplitude spatially mapped across the network. Amplitudes below a threshold of 0.1 are omitted to visualize pruning. Due to detrending, the relative time point of cut appears shifted to 19 min and 50 s. Scale bar = 1mm.

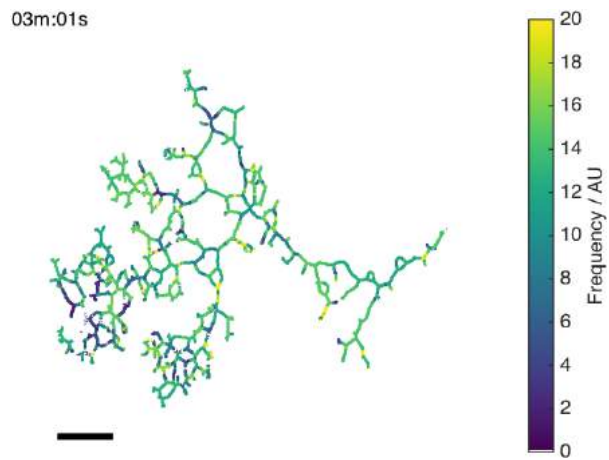

**Movie S3.** Spatially mapped oscillation frequency across the network. Mind that due to detrending, the relative time point of cut appears shifted to 19 min and 50 s. Scale bar = 1mm.

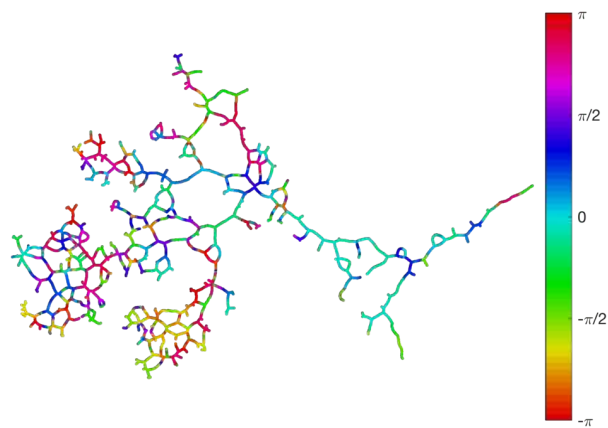

**Movie S4.** Spatially mapped phase pattern across the network with additional Gaussian smoothing in a 30px environment. Cut time is at 19 min and 50 s. Mind change in phase waves from before cut, during stalling and during correlated phase. The cut itself is not visible in the phase pattern due to low amplitude residual slime and spatial smoothing. It is recommended to follow a single colour over a time span to guide ones eye. The network spans maximal 8.4 mm along the x-axis.

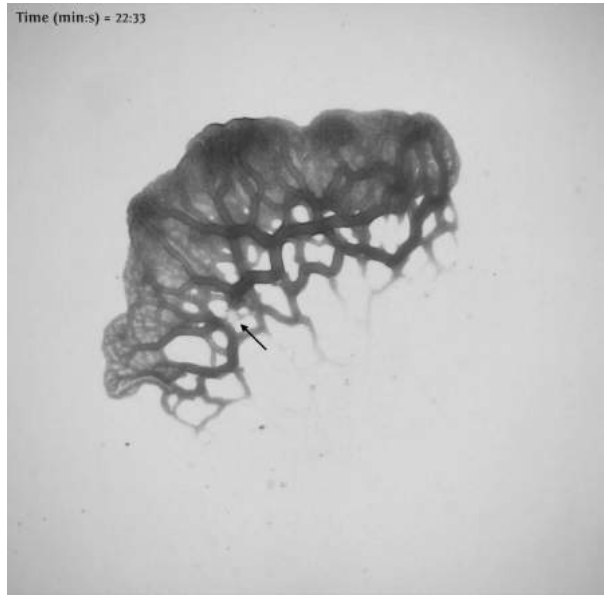

**Movie S5.** Bright field movie of E1 dataset showing severe stalling. Exemplary dataset for missing network morphology which makes network-based quantitative analysis impossible. Arrow marks cut location. Mind change in oscillation pattern, morphology and fan growth right after cut. Scale bar = 1mm.

**E1. List of experiments**

This section provides a complete list of acquired data sets.

Each experiment is a figure with two sub-figures displaying: a) the raw images of the plasmodium at the beginning of the experiment (left), immediately after the cut is performed (middle), and at the end of experiment (right). The cut site is marked with a red cross and the section of the network chosen for kymograph analysis is marked with a red line. The scale bar always represents 1mm and is shown in the latest image. b) shows the results of kymograph analysis. Here, the black bar, if applicable, denotes stalling.

The figures are complemented with a detailed descriptions of the morphological dynamics based on visual inspection of the bright field data. Both quantification and visual inspection together allow to identify if stalling does happen or not as discussed for each data set.

The experiments are sorted in a descending order by the magnitude of stalling. The organisms in experiments from Fig. E1 to Fig. E7 show most pronounced stalling, followed by a weaker display of stalling from Fig. E8 to Fig. E15. The organisms in experiments from Fig. E16 to Fig. E21 do not show a clear stalling behaviour.

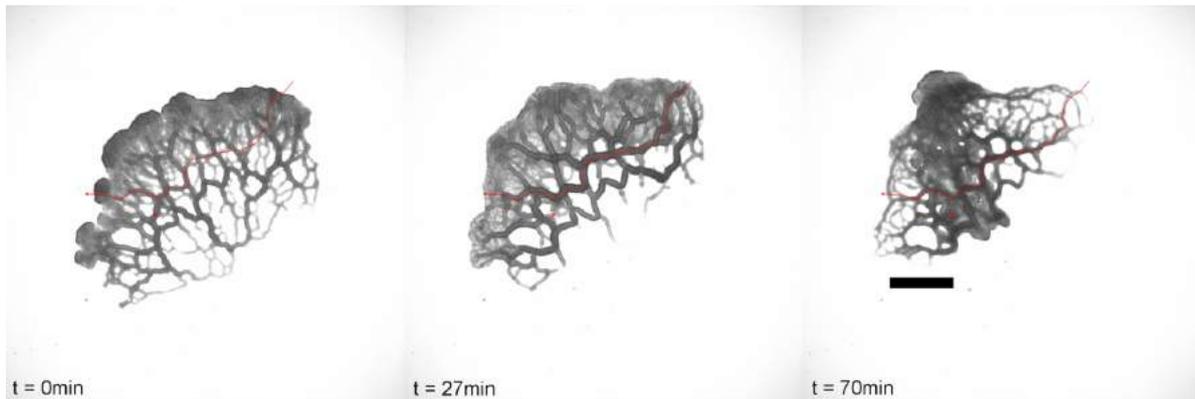

(a)

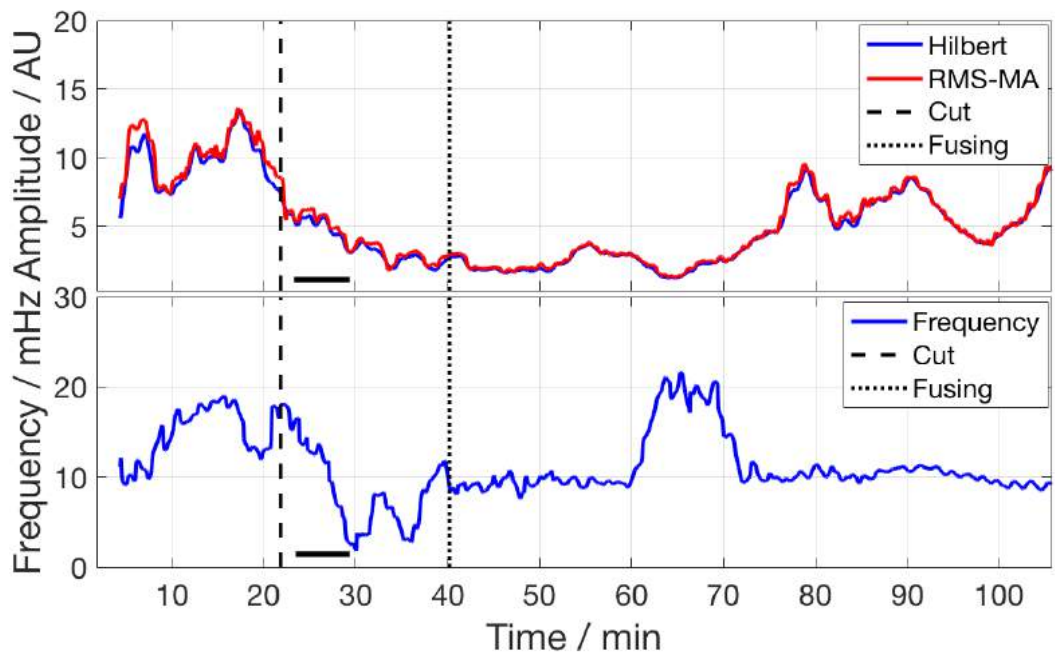

(b)

Figure E1: A small cut is performed across two of the thicker tube in this branched network, separating it into a large and small sub-network. A fan is first built around the cut site, then disintegrates as the flow is re-established. After an initial delay, the organism experiences stalling.

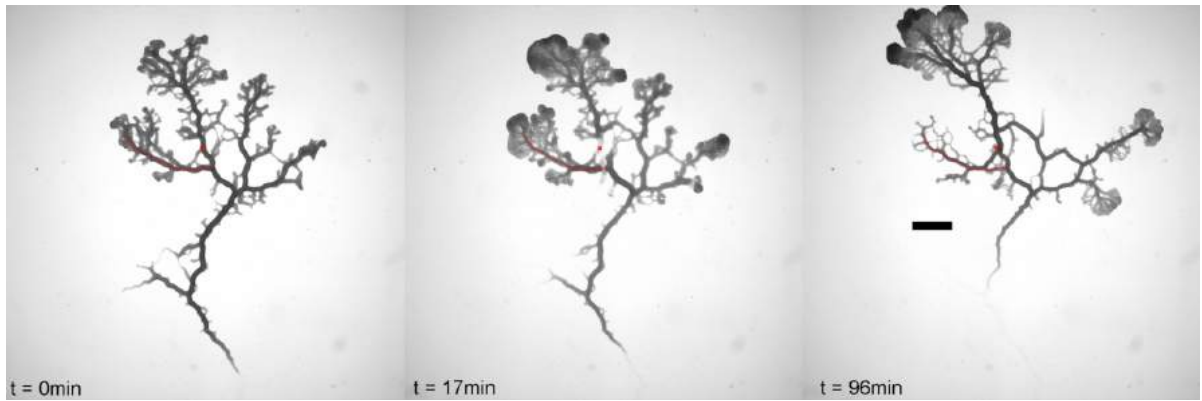

(a)

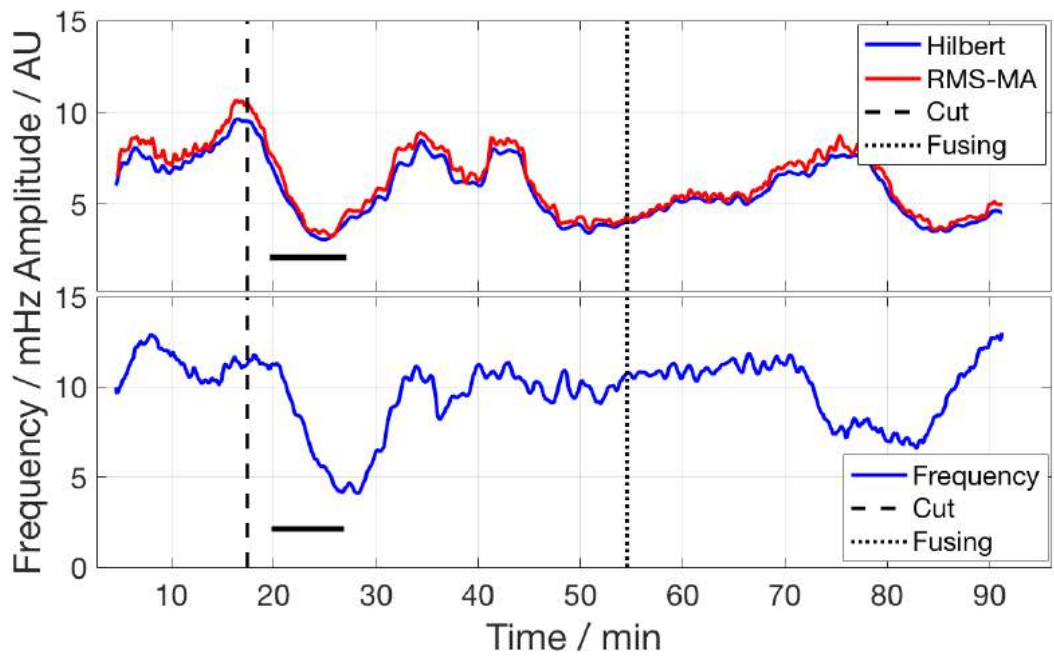

(b)

Figure E2: This data set is prominently described within the main text. We refer to the results section (3) for detailed description.

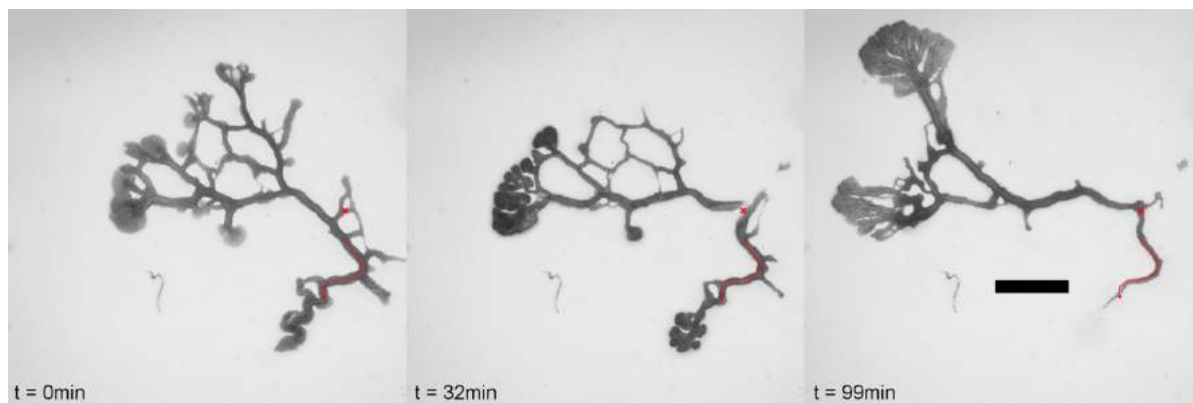

(a)

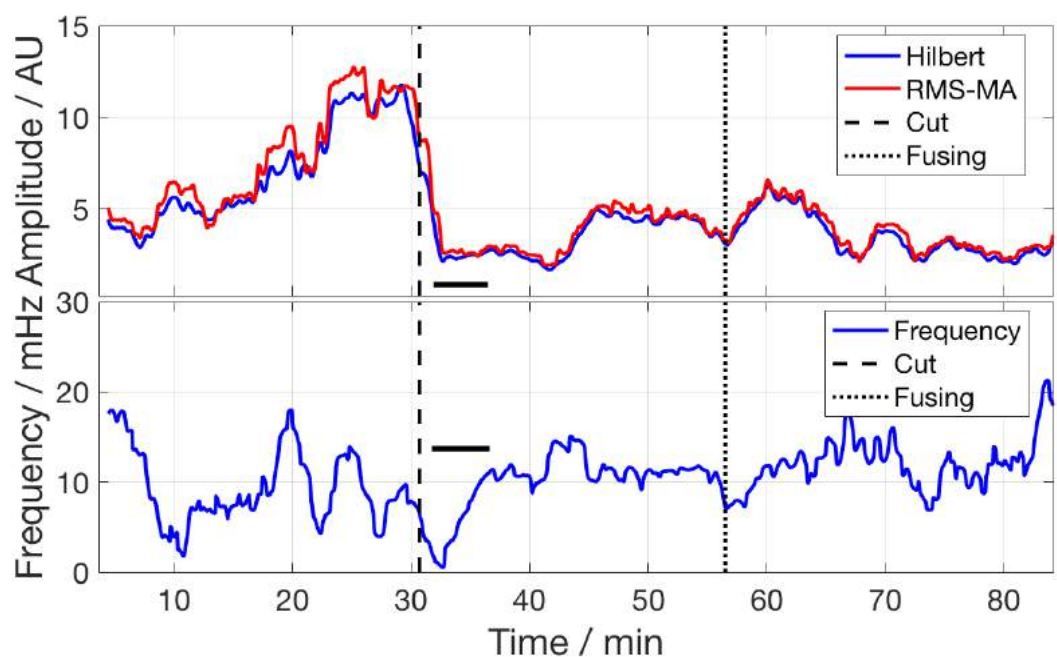

(b)

Figure E3: This small plasmodium with few tubes was cut across the thickest tube connecting the two fan regions. The cut removed a part of the tube and hence completely separated the two regions. Stalling is immediate after the cut, seen as a drop in amplitude and frequency. As the cut ends rejoin, a fan is built around the cut side. This experiment was fully analysed with quantitative analysis (see Fig.S5).

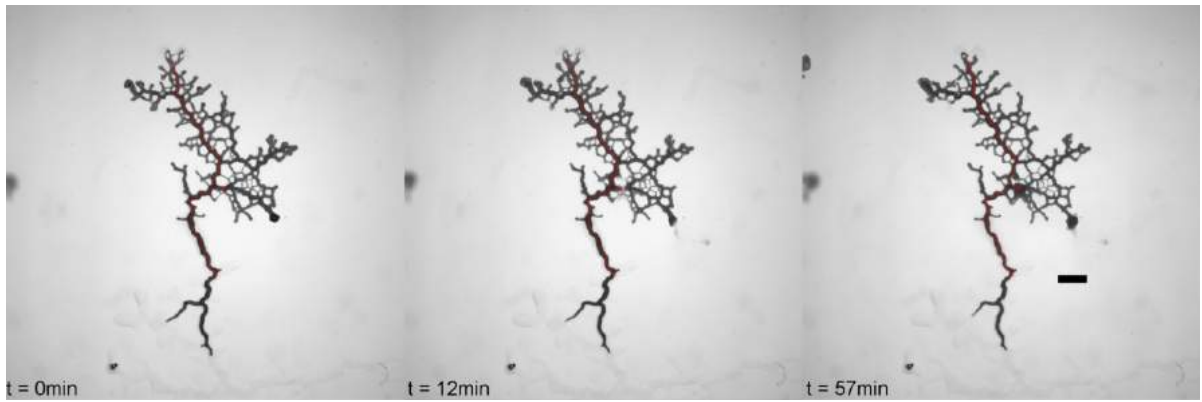

(a)

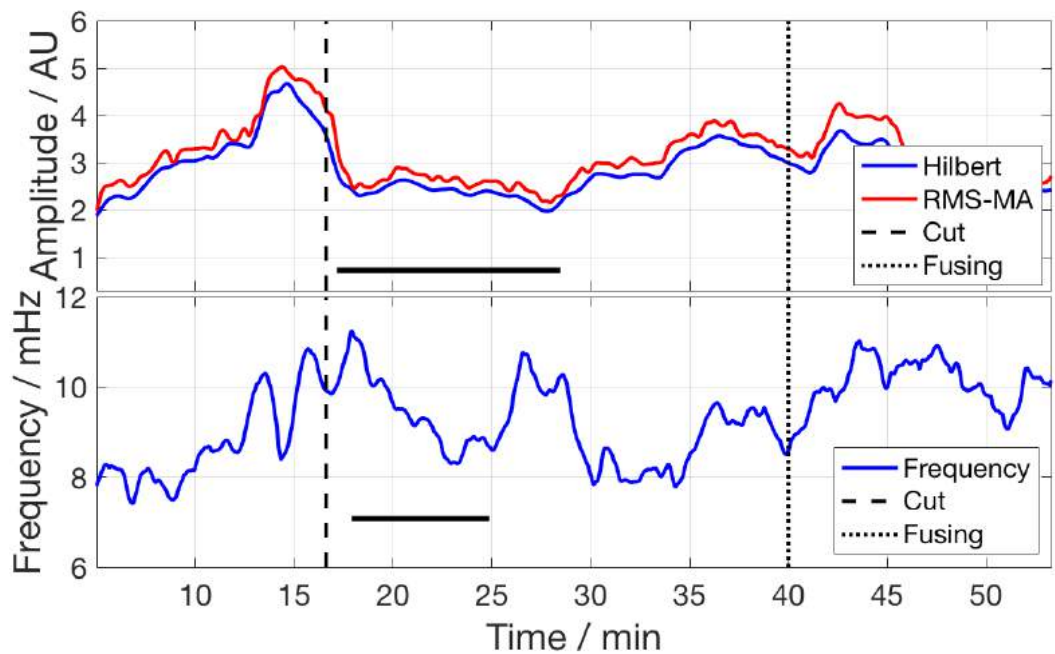

(b)

Figure E4: In this larger network, a thicker side tube was cut. The cut slightly separated the ends of the tube which then rejoin while a large fan is built around the cut site. The organism experiences stalling, visible as a decrease in frequency and amplitude of the oscillations. The fan around the cut site maintains its size until the end of the experiment.

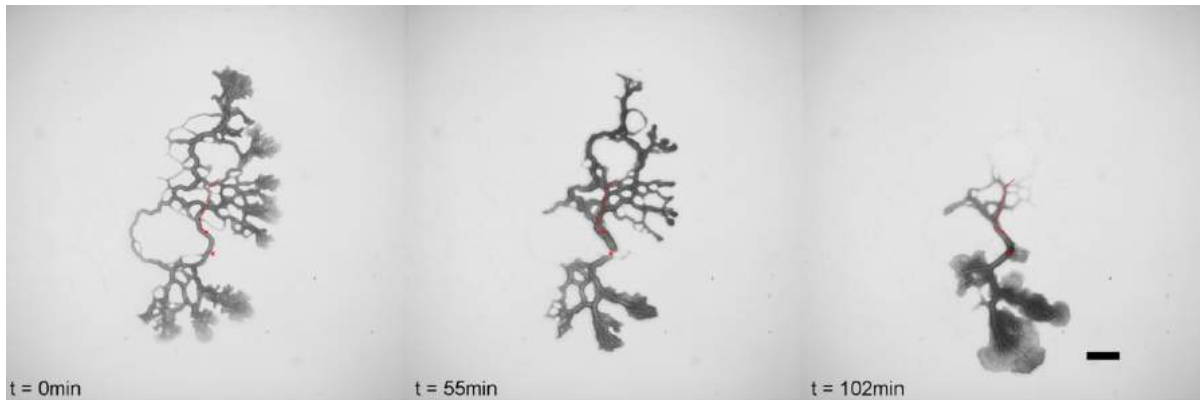

(a)

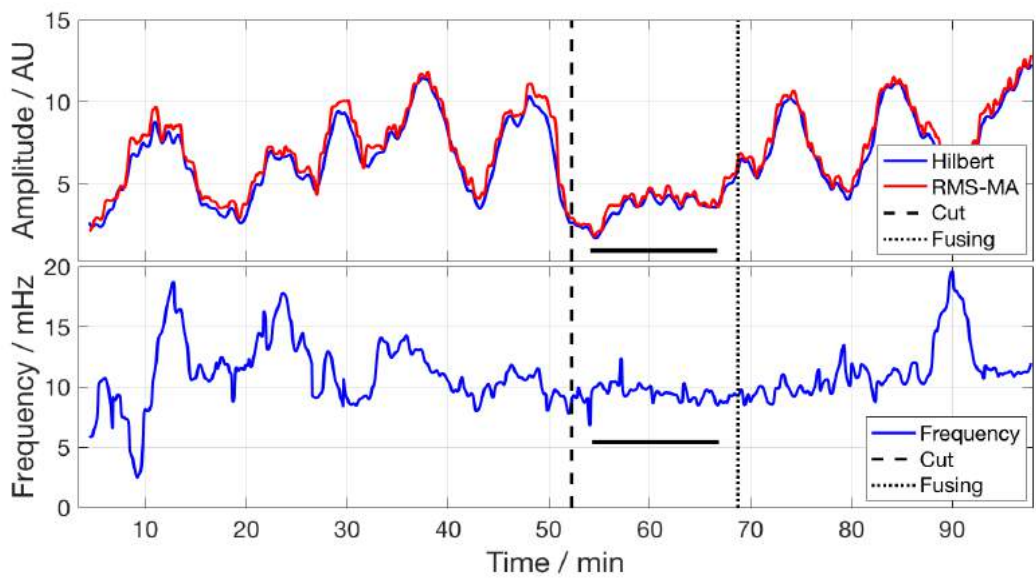

(b)

Figure E5: A cut was performed across a thick tube connecting two distant regions of the organism of comparable size, chosen such that the organism has no possibility of re-routing the flow through a close-by neighbouring tube. However, the cut barely separated the tube ends and they rejoin quickly. A fan is growing around the cut site. The stalling in both sub-networks appears in connection with very low variation both in amplitude and frequency of the oscillations compared to the pre-cut and post-rejoining phase.

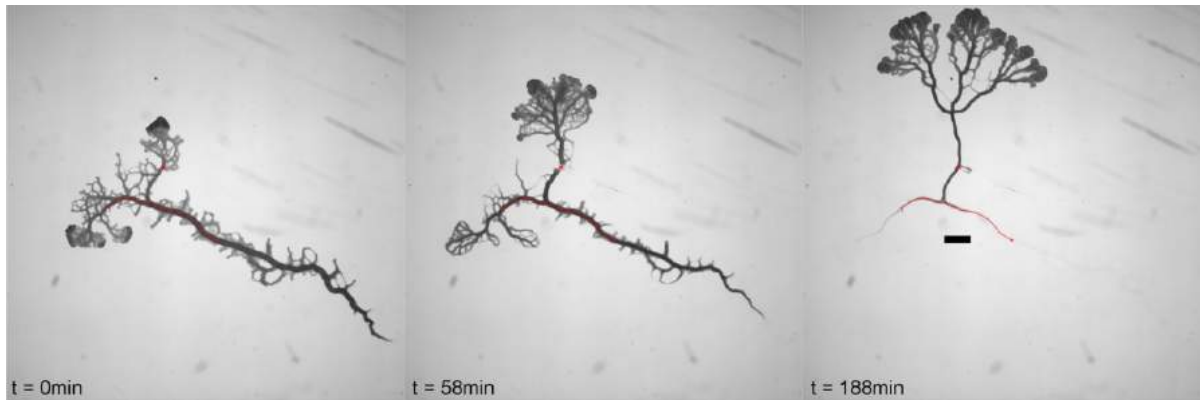

(a)

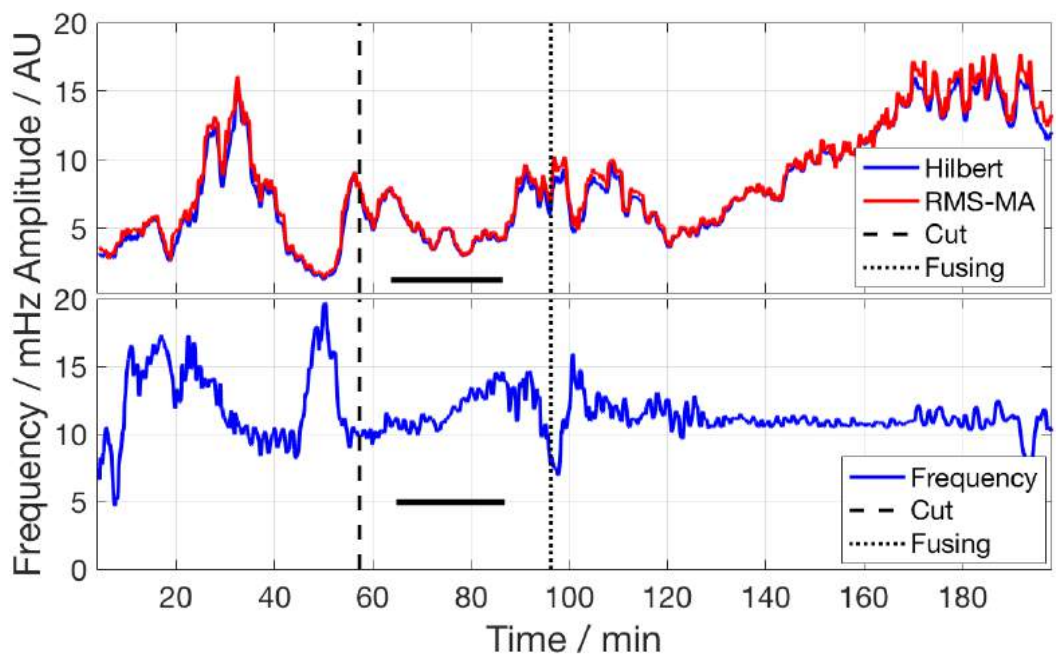

(b)

Figure E6: A cut is performed across a thick single tube connecting a large fan region with the rest of the organism. The amplitude and frequency of the oscillations stall first, followed by a gradual increase as the network rejoins the cut parts. Once the flow is re-established, the plasmodium prunes one side of the body and moves as a whole in the direction of the large fan.

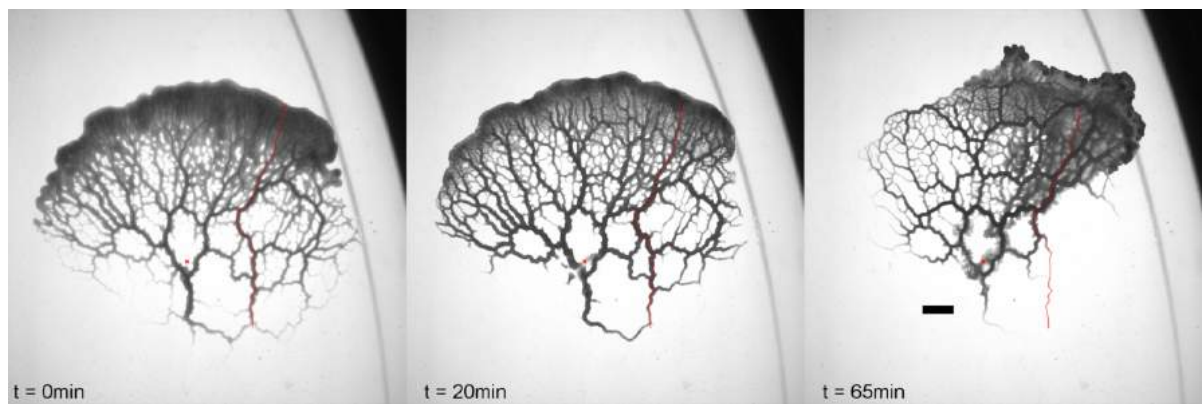

(a)

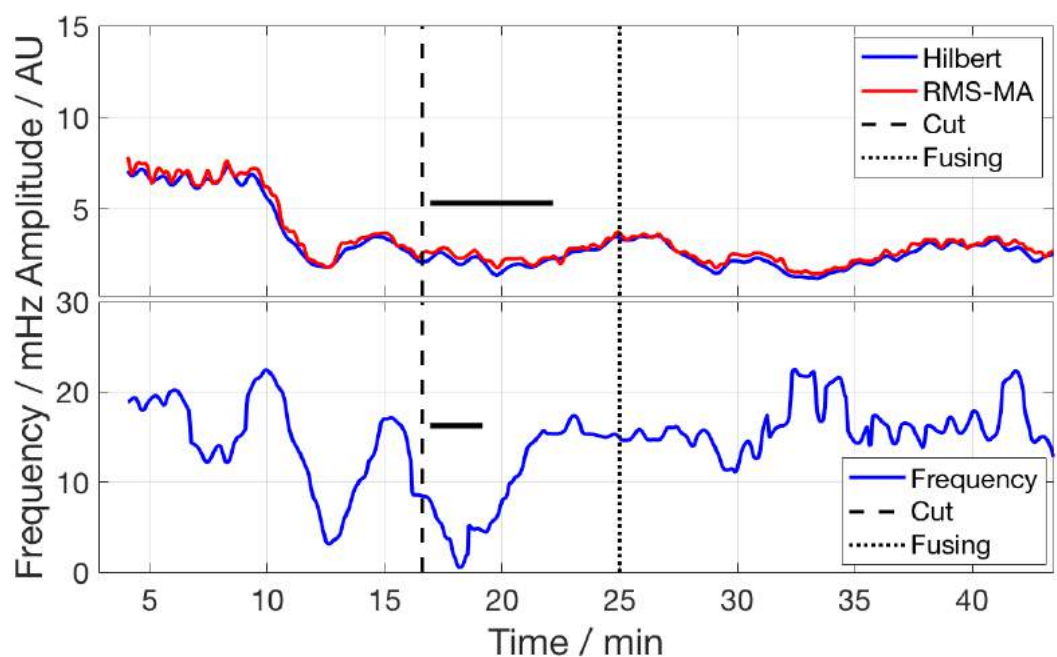

(b)

Figure E7: The morphology of this small network changes rapidly over the course of the experiment. A cut is performed across two neighbouring wide tubes at the base of the network. The tube segments fuse back completely as fans grow around it. Stalling is very prominent in this experiment and shows as a large drop in the frequency of the oscillations.

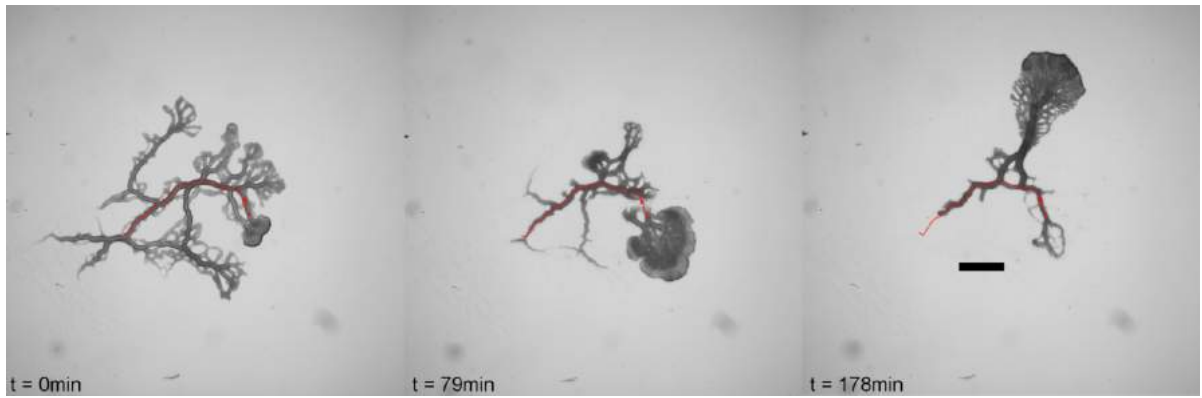

(a)

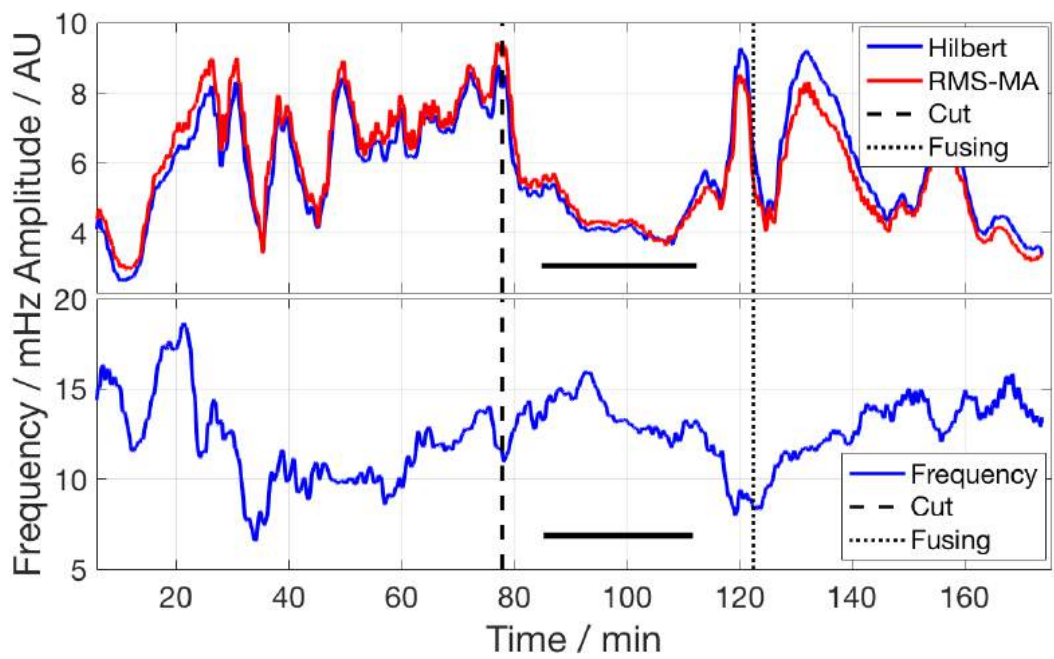

(b)

Figure E8: A cut is applied across the base of the only large fan in the network, resulting in damage of two thick tubes. The large fan, now separated, stalls and then builds body mass at the cut site. After rejoining, the organism as a whole builds a large fan and moves as a whole.

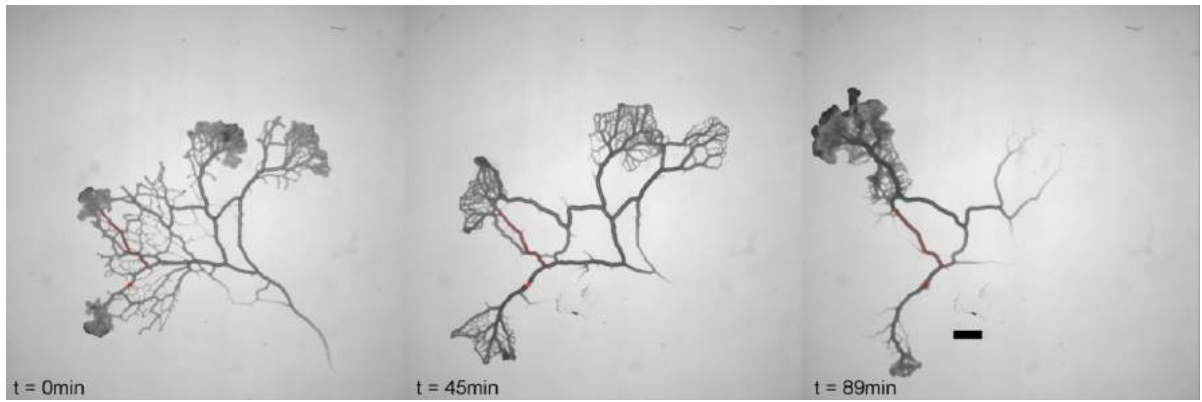

(a)

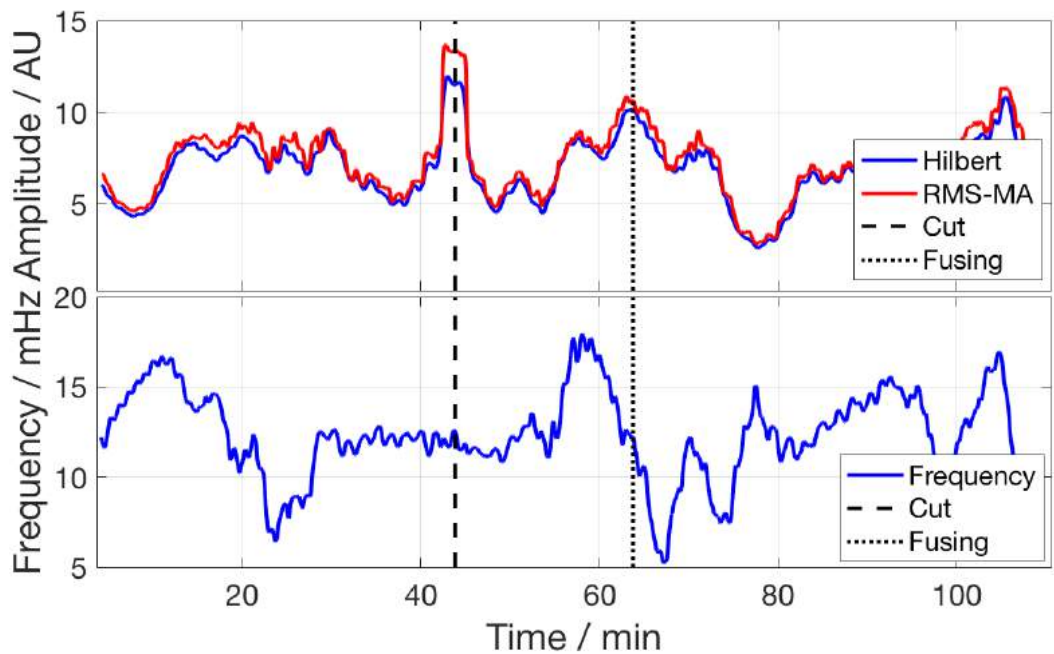

(b)

Figure E9: Here, four fan regions are initially growing and one of these is cut off at the only connection. The cut site is repaired fairly quickly with minimal fan growth. The amplitude and frequency hardly drop after the cut, yet fan growth in the periphery stalls and only resumes after flow is reinstated. After fusion the smaller sub-network is declining and fans are grown in the big sub-network.

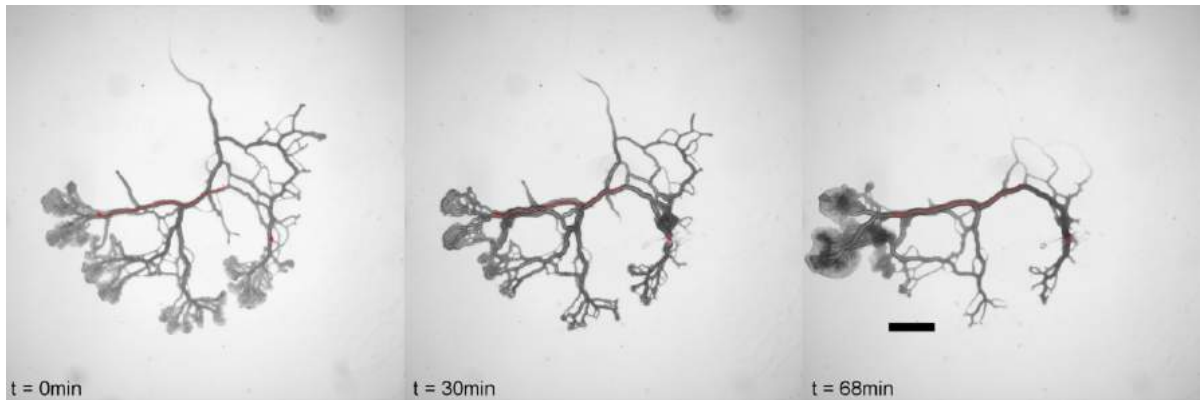

(a)

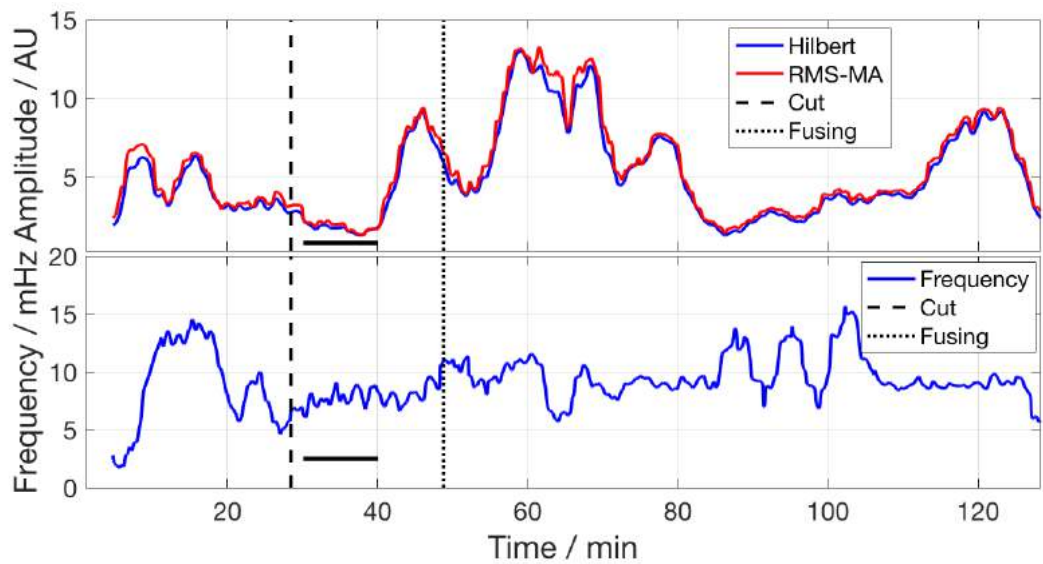

(b)

Figure E10: A cut is performed across one of the main tubes, distal from the region where most of the body mass is concentrated. The kymograph along the thickest tube shows stalling as a drop in the amplitude. A small fan grows at the site of the cut from both sub-networks. As the tubes refuse and the flow is established, the organism builds big fans and moves out of the imaging window.

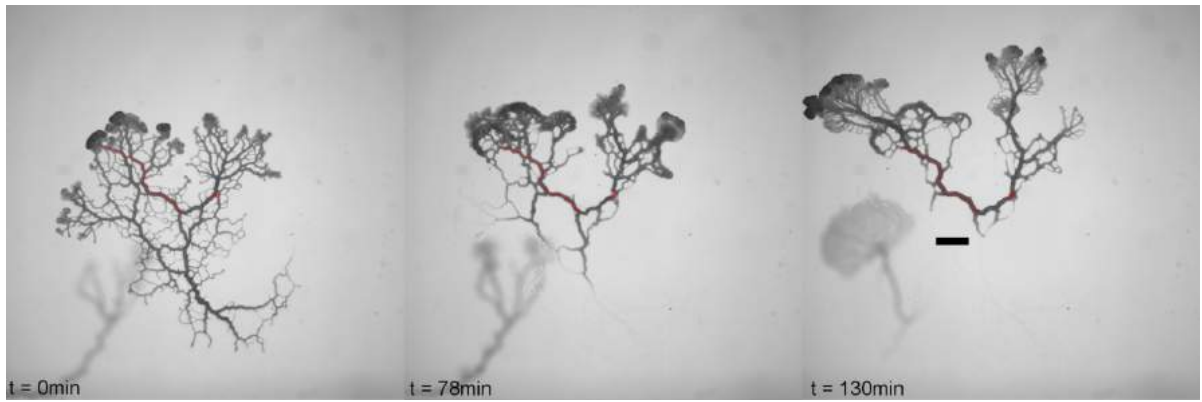

(a)

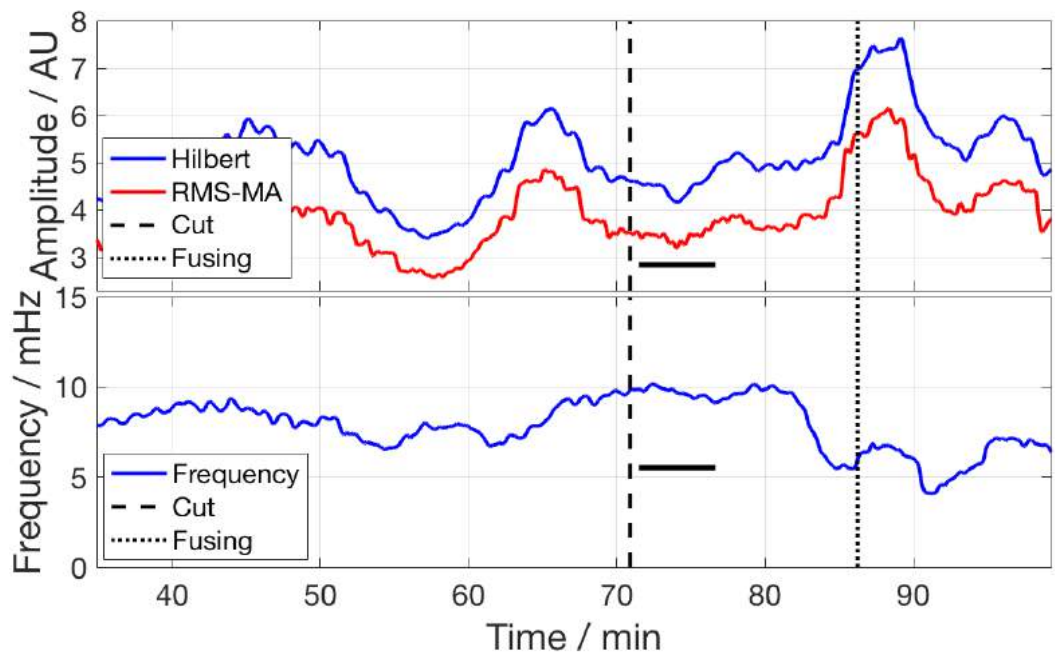

(b)

Figure E11: Initially the organism rapidly changes its morphology, likely as a response to light. A cut was performed across the biggest tube connecting the two fan regions, a site with no possibility of re-routing the flow through neighbouring tubes. The organism experiences slight stalling and quickly rejoins the cut tube ends. A small fan is built around the cut site.

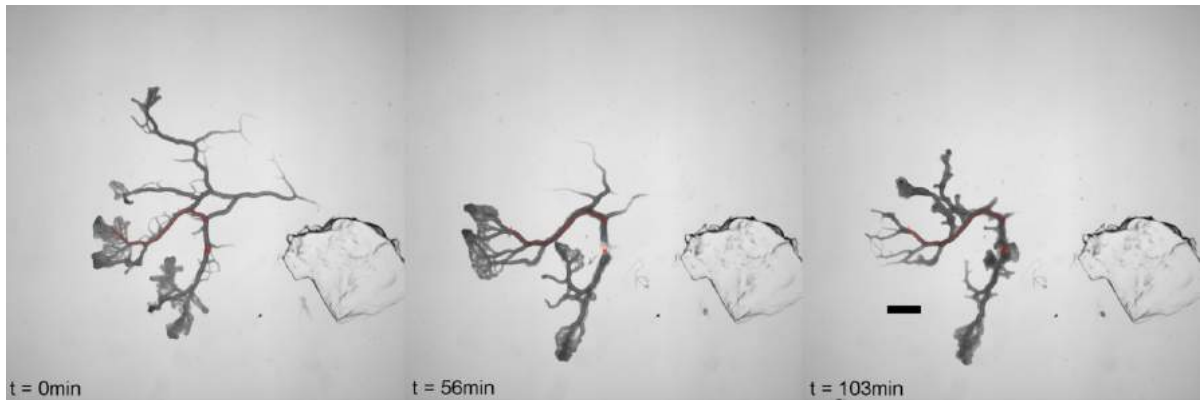

(a)

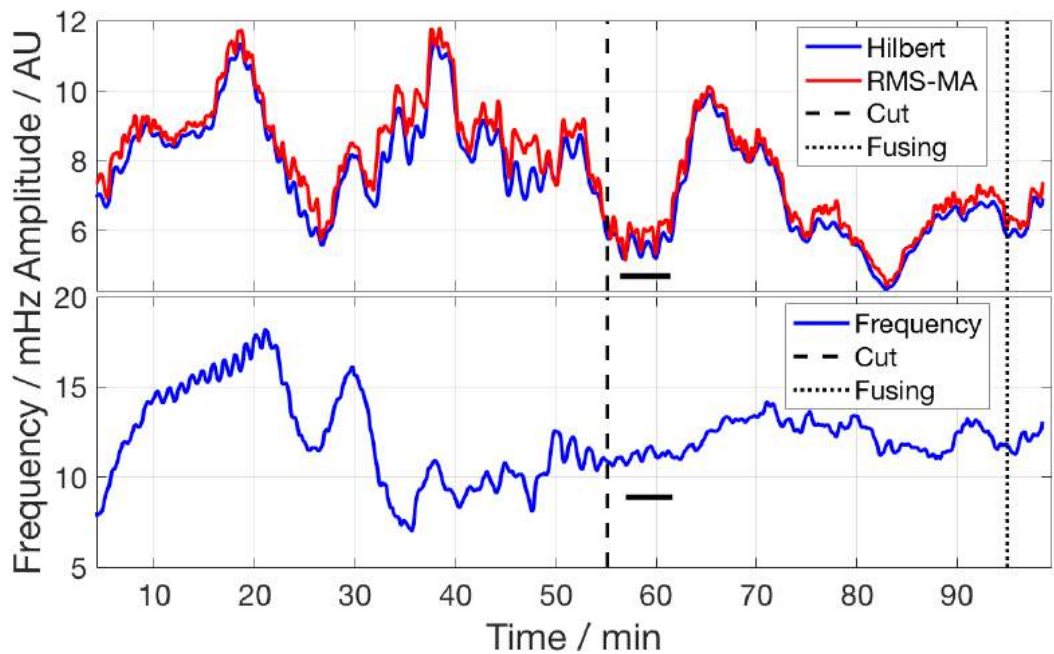

(b)

Figure E12: The organism experiences a putative light shock in the beginning indicated by the change in morphology. The cut is applied across a thick tube connecting two regions of comparable size and shape, separating them completely. The two sub-networks take a relatively long time to rejoin, meanwhile amassing new thick fans out of existing veins. Stalling is visible as a drop in amplitude. The general behaviour of this network suggests that the effects of the initial light shock persist throughout the whole experiment.

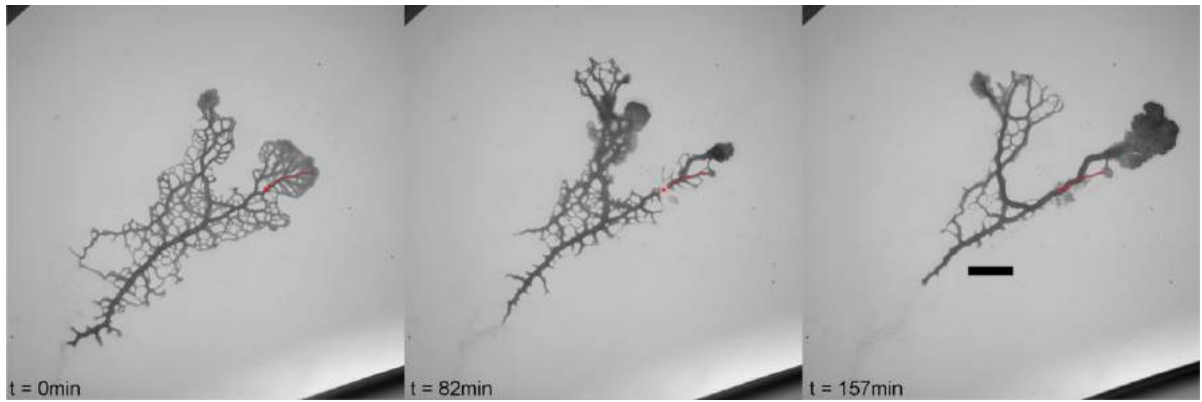

(a)

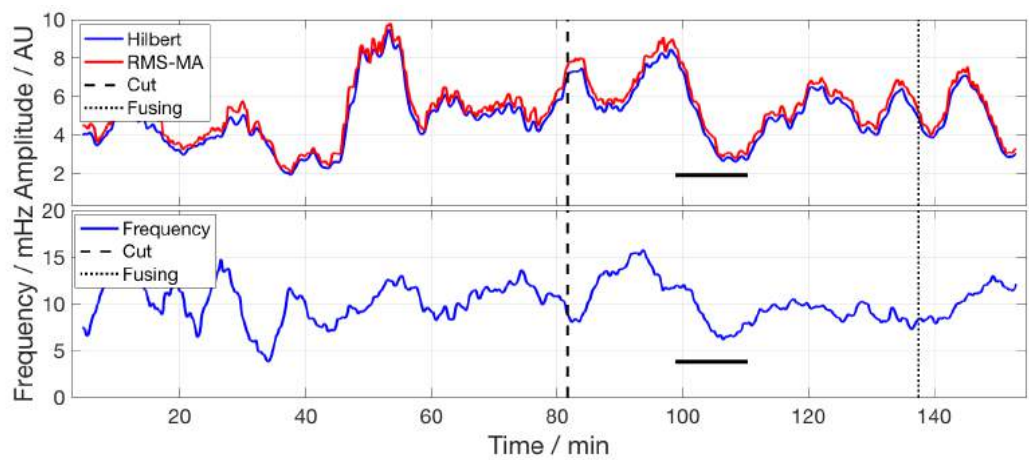

(b)

Figure E13: The cut is applied at the base of a smaller fan, followed by movement of the body mass in the fan towards the cut site. The fan in the larger sub-network retracts and the body mass moves towards the cut site. After the cut site is repaired and the sub-networks rejoin, the organism moves as a whole. The stalling occurs in both networks after the cut with a delay.

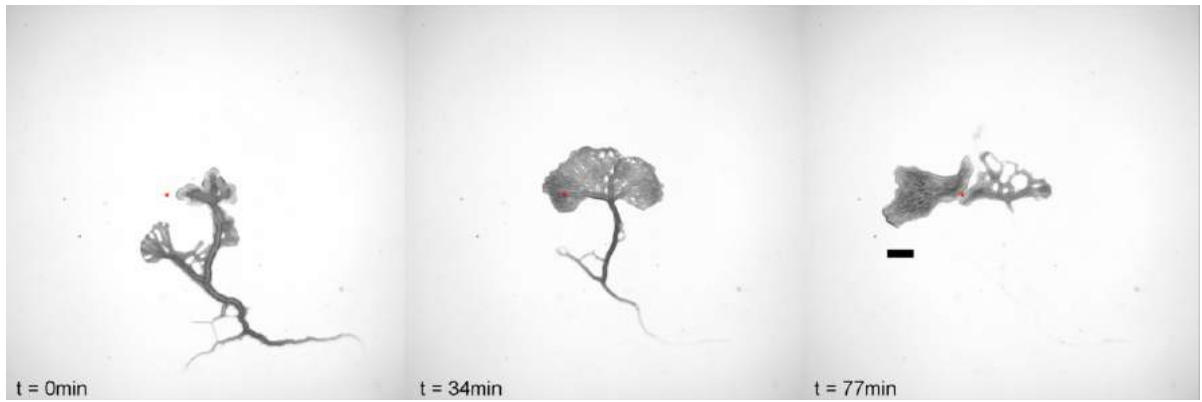

(a)

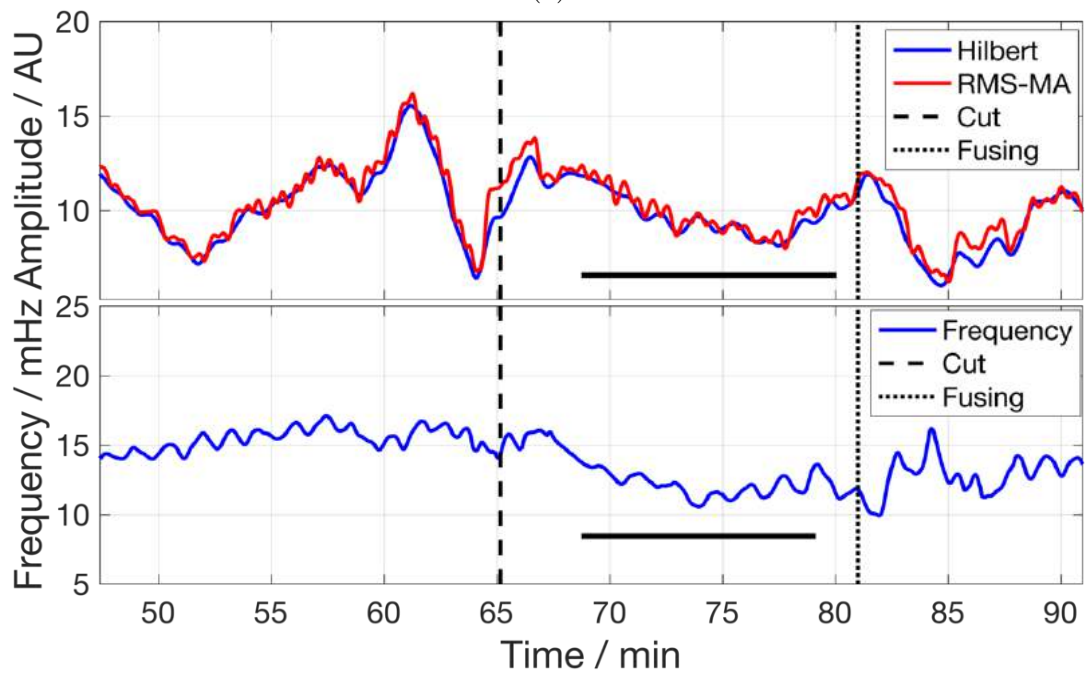

(b)

Figure E14: This plasmodium behaves rather like a very motile foraging fan than an extended network. There are hardly any hard distinctions between tube and fan in the main part of the plasmodium. The cut completely severs the specimen into two parts and stalling is evident, but it quickly returns to foraging behaviour.

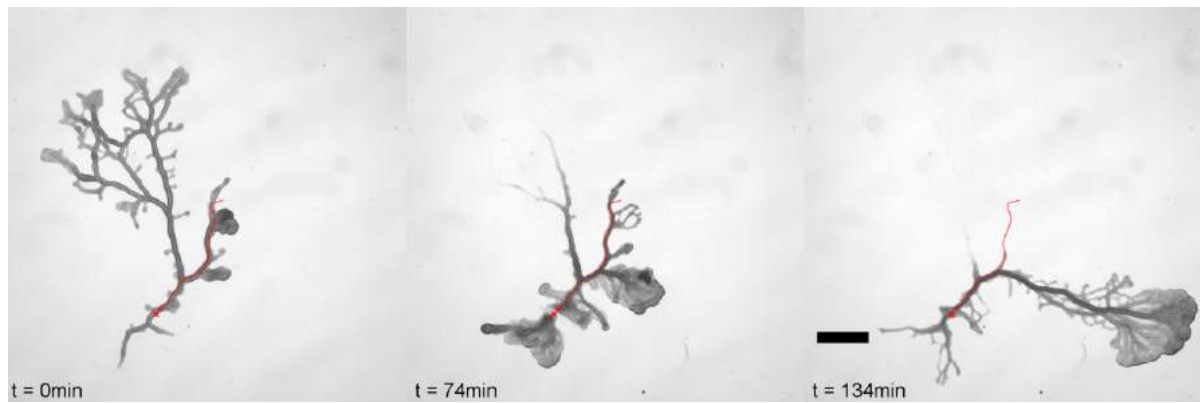

(a)

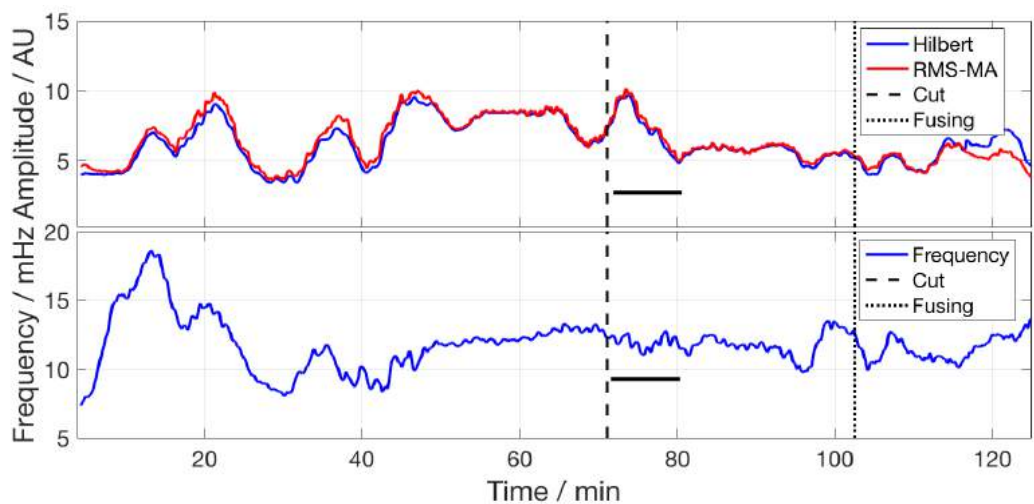

(b)

Figure E15: The network in this experiment initially shows a fast morphology change, possibly from illumination, and fan growth out of the sides of the tubes. A cut is performed along the base of a fan. The network experiences hardly any stalling, with a slight decrease in amplitude and frequency of the oscillations. The cut parts rejoin, the organism builds a large fan and moves in its direction. Is it possible that the initial effects persist during the whole experiment.

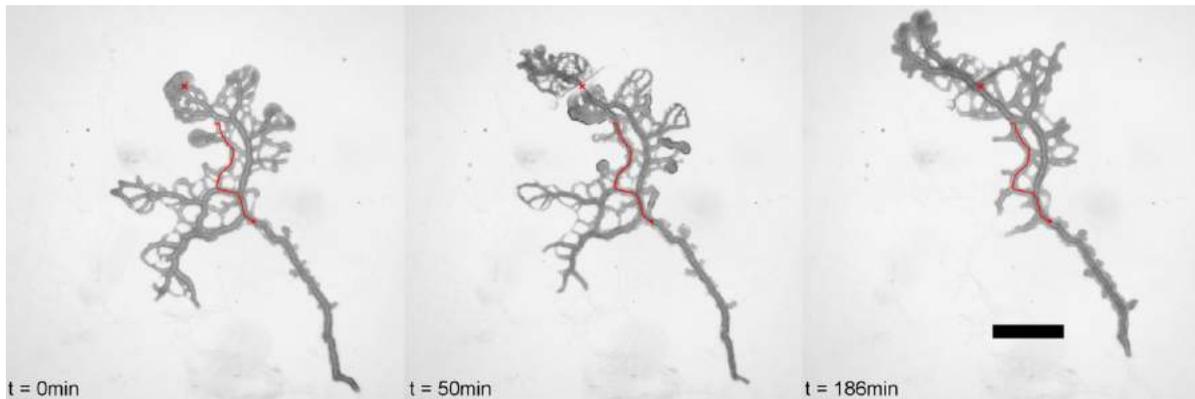

(a)

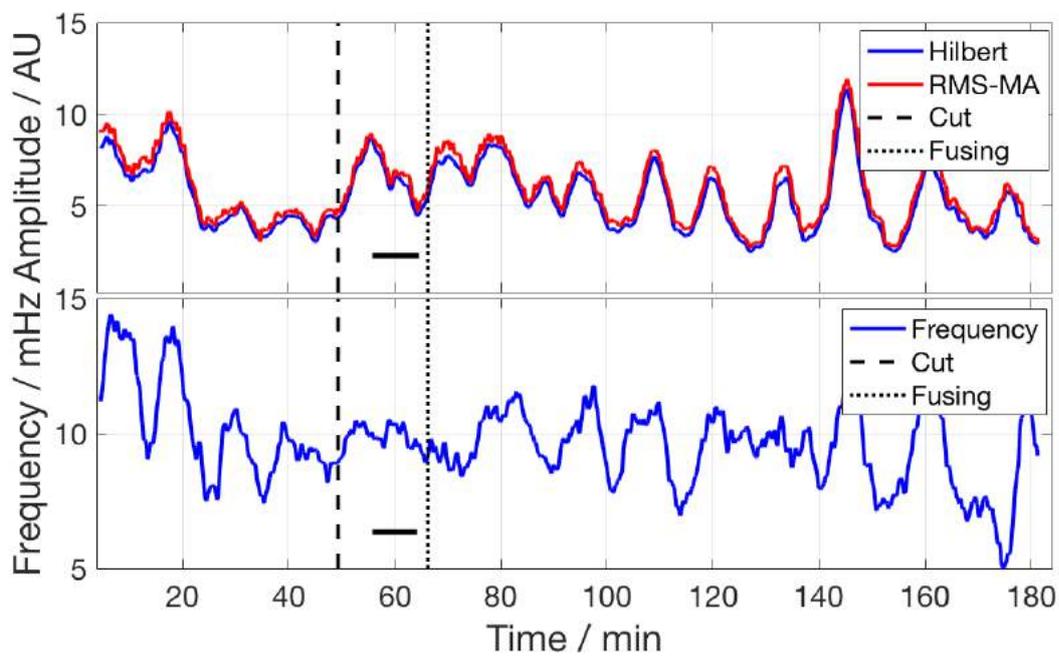

(b)

Figure E16: The plasmodium experiences a putative initial light shock which is indicated by a quick change of morphology, resulting in a network with a single fan. A cut was performed across the tube in the fan. The cut parts rejoin quickly. A thick fan is first grown at the cut site, then absorbed into the fan at the top of the image. No pronounced stalling is visible.

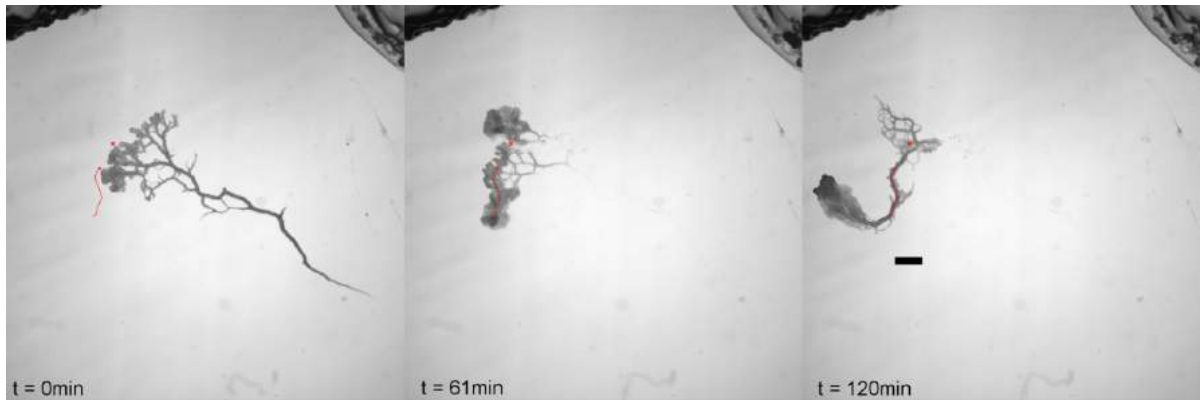

(a)

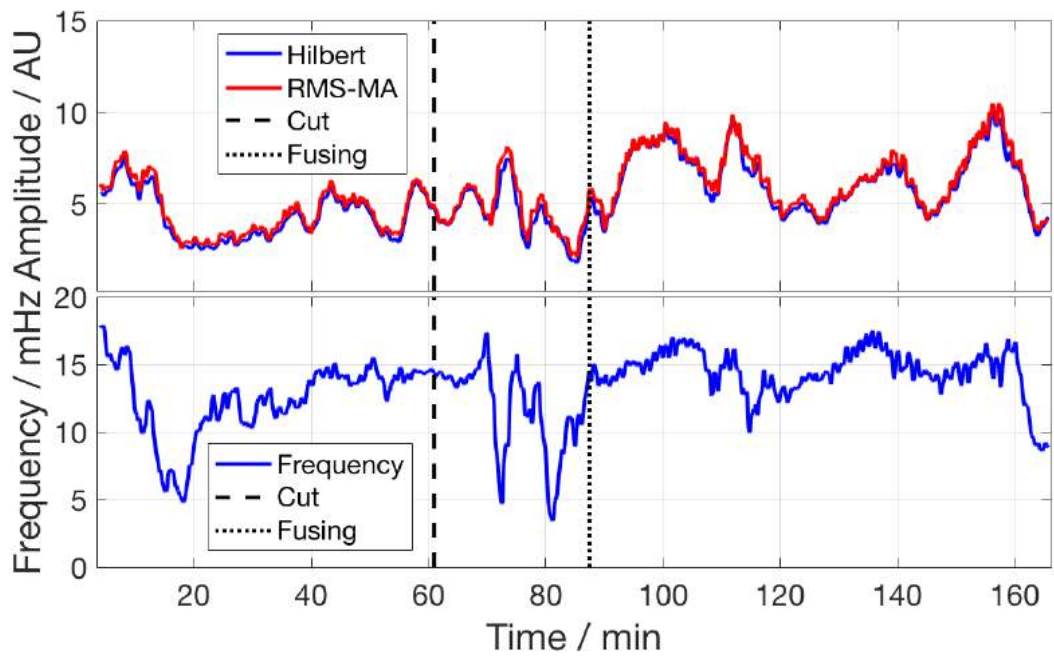

(b)

Figure E17: The plasmodium has high motility in this experiment. A cut is performed across a tube connecting two fan regions. After the cut, the material from the fan closer to the cut is moved towards the cut site and the cut parts rejoin. There is no obvious stalling.

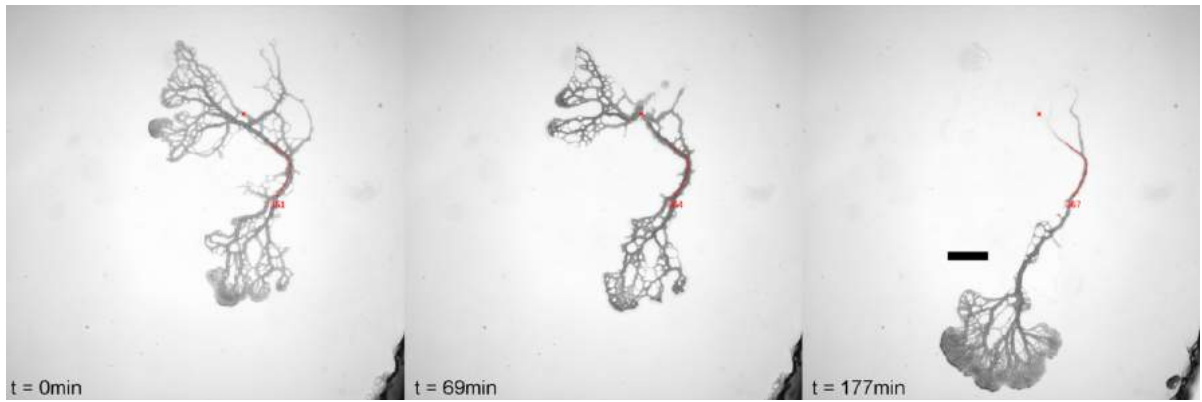

(a)

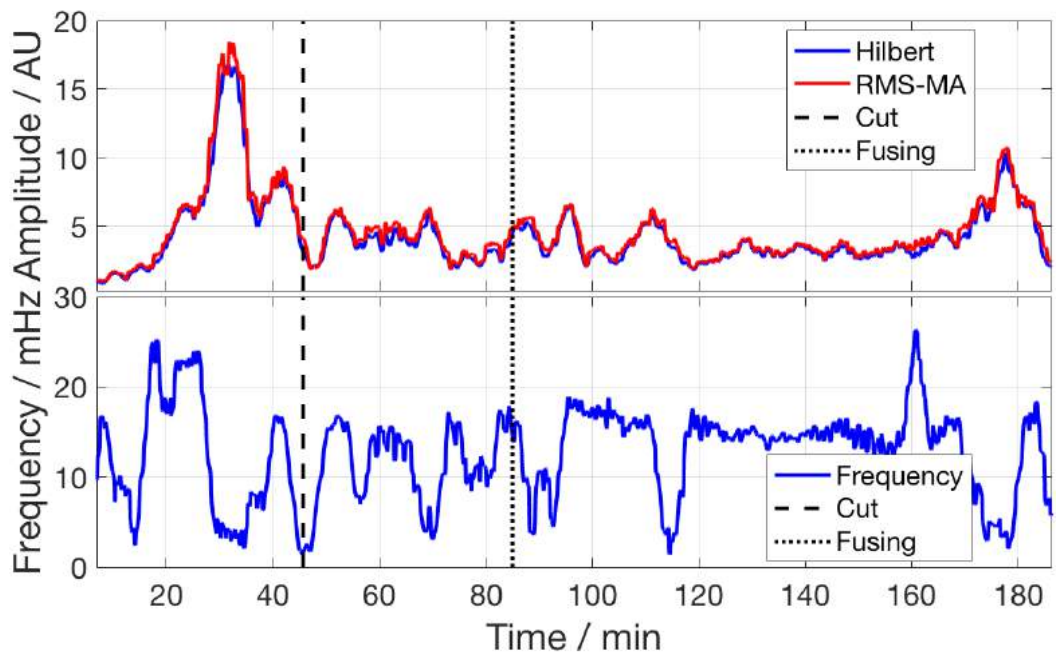

(b)

Figure E18: A cut is made across the thickest tube connecting two fan regions at the opposing ends of the organism. The organism builds a large, diffuse fan at the cut site and rejoins the cut parts. The network then moves as a whole. The variations in the frequency and amplitude of the osculations make the identification of stalling challenging.

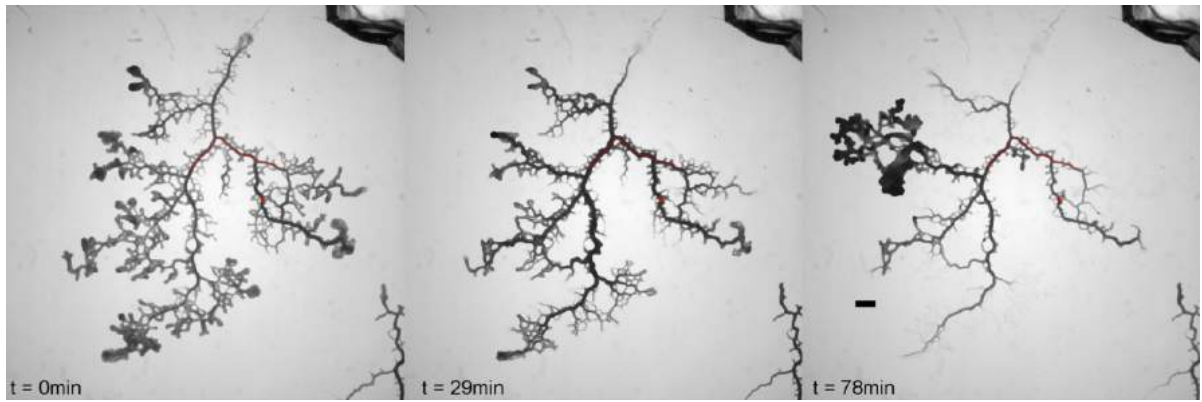

(a)

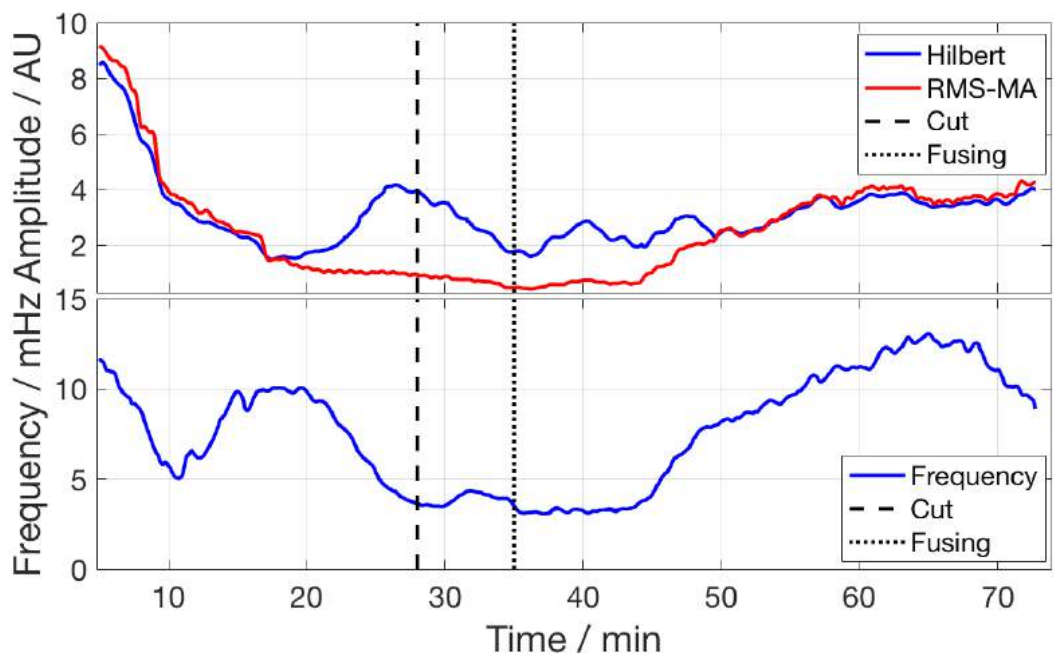

(b)

Figure E19: In this large branched network, a cut is applied across on of the thick veins, damaging it slightly. The cut parts re-join quickly and no stalling is observed.

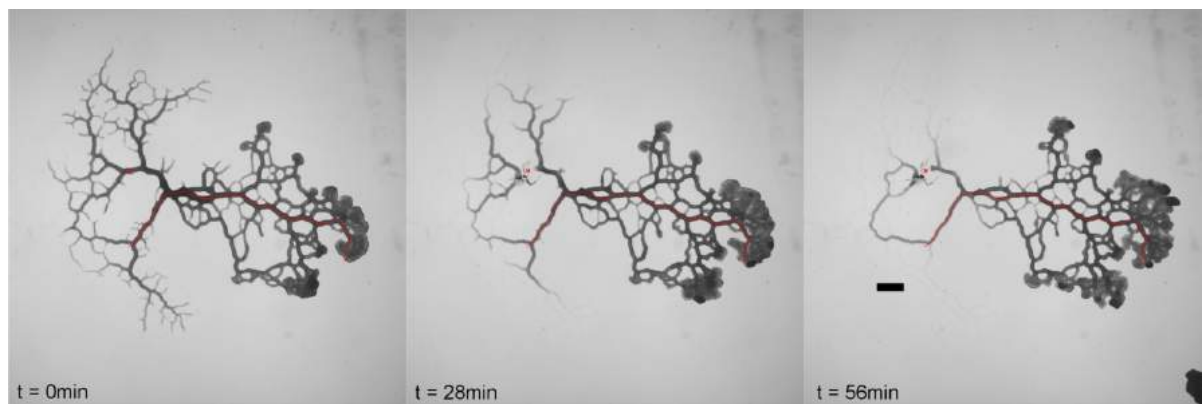

(a)

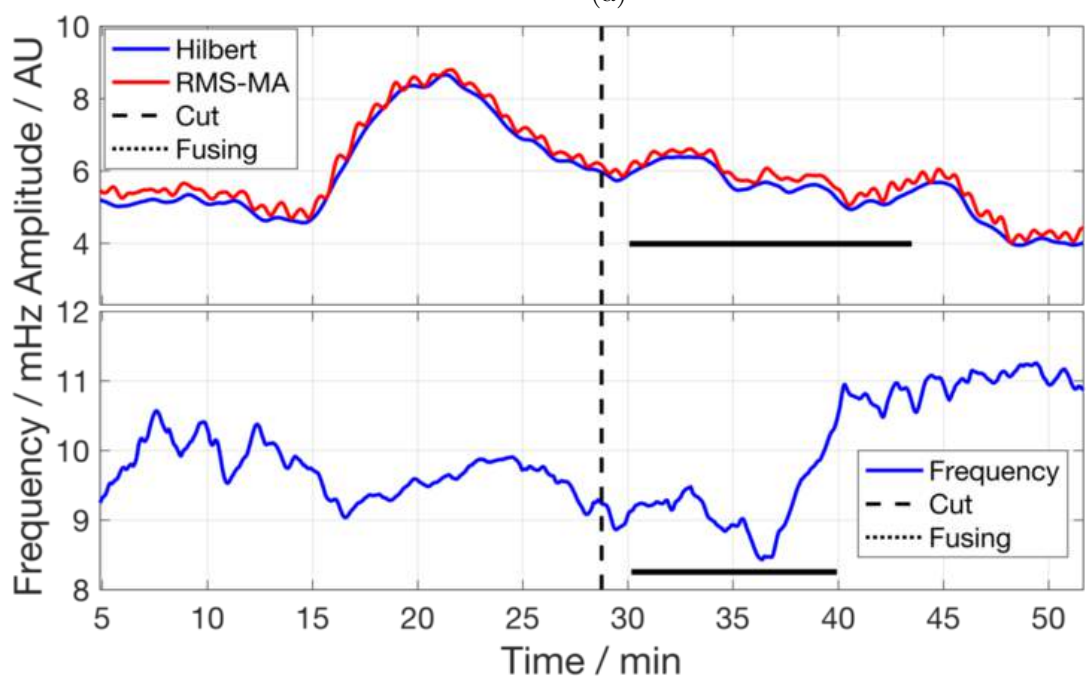

(b)

Figure E20: Here a section in the trailing part of the plasmodium is cut. Fan growth does occur, but hardly closes the gap as the neighbouring tubes are pruning to continue foraging in the distal fan region. Stalling is hardly noticeable and not evident from the amplitude graph.

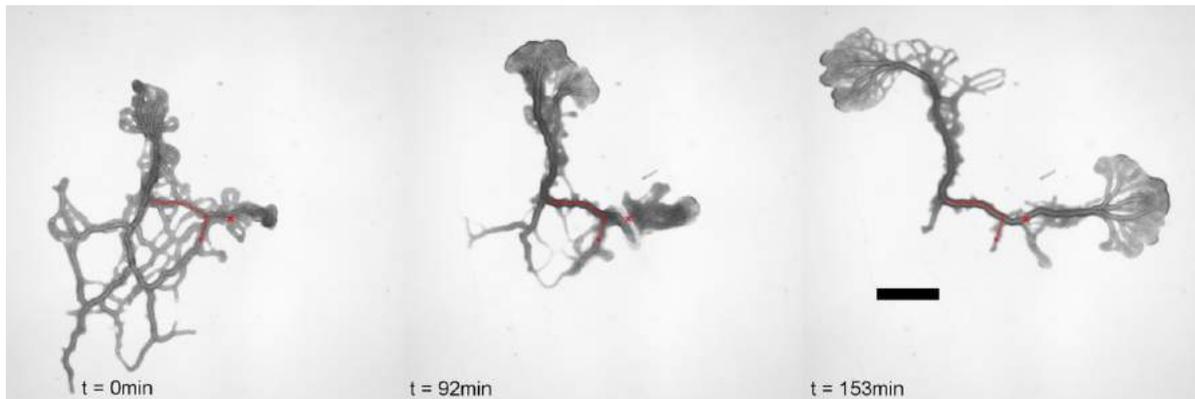

(a)

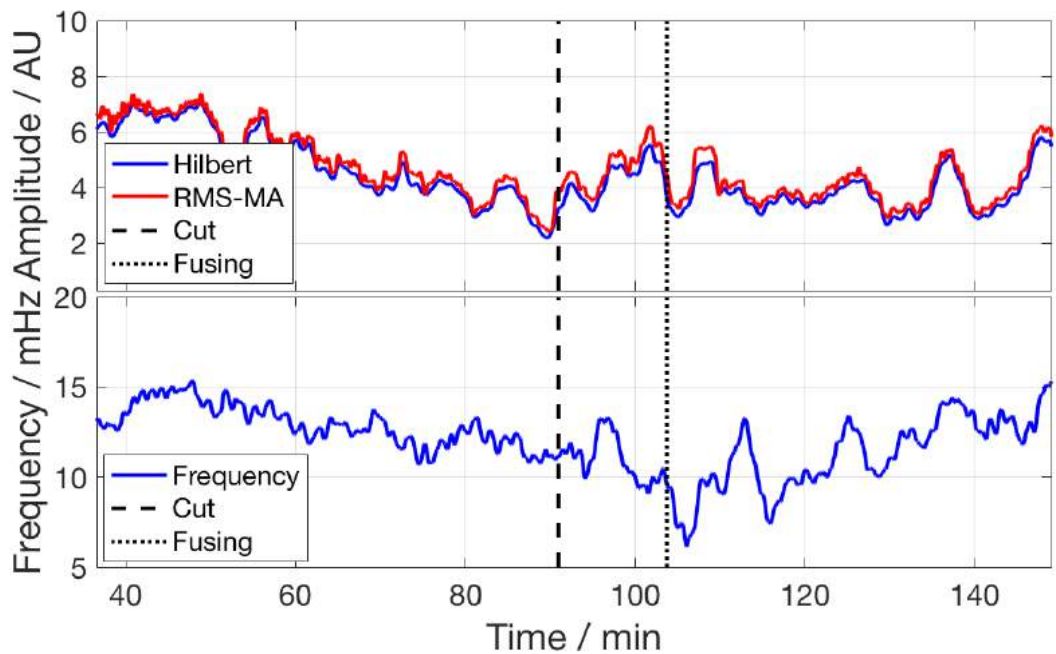

(b)

Figure E21: The small plasmodium in this experiment first undergoes a likely reaction to light, recovering into a structure made out of two fans connected by a thick tube. A cut is performed across the base of one of the fans. The tube, as well as the plasmodial sheet around it, rejoin quickly and the flow is re-established. There is no apparent stalling.


\end{document}